\documentclass[journal, onecolumn,12pt, draftclsnofoot]{IEEEtran}
\usepackage{subfigure}
\usepackage{citesort}
\usepackage{amsmath,amsthm}
\usepackage{algorithmic,algorithm}
\usepackage{array}
\usepackage{amsfonts}
\usepackage{graphicx,color,overpic,psfrag,epsfig}
\usepackage{bm}
\usepackage{footmisc}
\usepackage{xcolor}
\usepackage{epstopdf}
\usepackage{amssymb}%the square at the end of proof

\newtheorem{proposition}{Proposition}

\newtheorem{lemma}{Lemma}

\newcommand{\qa}{{\bf a}}
\newcommand{\qg}{{\bf g}}
\newcommand{\qh}{{\bf h}}
\newcommand{\qn}{{\bf n}}
\newcommand{\qp}{{\bf p}}
\newcommand{\qu}{{\bf u}}
\newcommand{\qv}{{\bf v}}
\newcommand{\qx}{{\bf x}}
\newcommand{\qy}{{\bf y}}

\newcommand{\qA}{{\bf A}}

\newcommand{\qG}{{\bf G}}
\newcommand{\qH}{{\bf H}}
\newcommand{\qI}{{\bf I}}

\newcommand{\qP}{{\bf P}}

\newcommand{\qU}{{\bf U}}

\newcommand{\bbC}{{\mathbb C}}
\newcommand{\calCN}{{\mathcal {CN}}}

\newcommand{\calS}{{\mathcal S}}

\newcommand{\tr}{{\sf tr}}
\newcommand{\qzero}{{\bf 0}}
\newcommand{\Ex}{{\sf E}}

\newcommand{\argmax}{\operatornamewithlimits{arg\, max}}

\begin{document}

\title{Joint Optimization of Fronthaul Compression and Bandwidth Allocation in Uplink H-CRAN with Large System Analysis}

%\author{\IEEEauthorblockN{Wenchao Xia$\stackrel{\dag}{,}$ Jun Zhang$\stackrel{\dag}{,}$ Tony Q. S. Quek $\stackrel{\P}{,}$ Shi Jin$\stackrel{\ddag}{,}$ and Hongbo Zhu$\stackrel{\dag}{}$}}

\author{Wenchao Xia\thanks{W. Xia, J. Zhang, and H. Zhu are with the Jiangsu Key Laboratory of Wireless Communications, Nanjing University of Posts and Telecommunications, Nanjing 210003, P. R. China, E-mail addresses: {\sf \{ 2015010203,zhangjun,hbz\}@njupt.edu.cn}.}, Jun Zhang, Tony Q. S. Quek\thanks{T. Q. S. Quek is with the Singapore University of Technology and Design, Singapore 487372, E-mail address: {\sf tonyquek@sutd.edu.sg}.}, Shi Jin\thanks{S. Jin is with the National Mobile Communications Research Laboratory, Southeast University, Nanjing 210096, P. R. China, E-mail addresses: {\sf jinshi@seu.edu.cn}.}, and Hongbo Zhu \thanks{This work was presented in part at IEEE Global Communications Conference (GLOBECOM) \cite{Xia2017Joint}, Singapore, Dec. 2017.}
}

\maketitle
\vspace{-1cm}
\begin{abstract}
In this paper, we consider an uplink heterogeneous cloud radio access network (H-CRAN),  where a macro  base station (BS) coexists with many remote radio heads (RRHs). For cost-savings, only the BS is connected to the baseband unit (BBU) pool via fiber links. The RRHs, however, are associated with the BBU pool through wireless fronthaul links, which share the spectrum resource with radio access networks. Due to the limited capacity of fronthaul, the compress-and-forward scheme is employed, such as point-to-point compression or  Wyner-Ziv coding. Different decoding strategies are also considered.  This work aims to maximize the uplink ergodic sum-rate (SR) by jointly optimizing quantization noise matrix and bandwidth allocation between radio access networks and fronthaul links, which is a mixed time-scale issue. To reduce computational complexity and communication overhead, we introduce an approximation problem of the  joint optimization problem based on large-dimensional random matrix theory, which is a slow time-scale issue because it only depends on statistical channel information. Finally, an algorithm based on Dinkelbach's algorithm is proposed to find the optimal solution to the approximate problem. In summary, this work provides an economic solution to the challenge of constrained fronthaul capacity, and also provides a framework with less computational complexity to study how bandwidth allocation  and fronthaul compression can affect the SR maximization problem.
\end{abstract}
\vspace{-0.5cm}
\begin{IEEEkeywords}
 H-CRAN, fronthaul compression, compress-and-forward, Wyner-Ziv coding, point-to-point compression, bandwidth allocation, random matrix theory.
\end{IEEEkeywords}

\section{Introduction}
With the development of the Internet of things and mobile communication networks, the demands for high-speed data applications are growing exponentially recently \cite{Peng2015System}. Meeting such challenging goals should involve new system architectures  and advanced signal processing for wireless communications \cite{tony2017cloud}.
The paradigm of heterogeneous networks (HetNets) \cite{Gao2018throyghput}, composed of a hierarchy of  macro cells enhanced by small cells of different sizes, has attracted lots of attention from both industry and academia.  Macro cells with high-power base stations (BSs) provide ubiquitous coverage and  small cells use various radio access technologies to serve user equipment terminals (UEs) with high data-rate demands. Unfortunately, the inter- and intra-tier interferences, resulting from densification of small cells, restrict the improvement of performance gains and commercial applications of HetNets.

At the same time, cloud computing has emerged as  a  popular  computing  paradigm for enhancing both spectral and energy efficiencies \cite{Tang2017System}. As an application of cloud computing to radio access networks, cloud radio access networks (C-RANs) have been proposed to achieve cooperative gains.  In C-RANs, radio frequency processing is implemented at remote radio heads (RRHs) whereas baseband processing is centralized in a baseband unit (BBU) pool. However, the performance improvement of C-RANs is restricted by capacity-limited fronthaul links. Furthermore, since C-RANs are mainly used in hotspots to provide high data rates, control signalling and real-time voice service are not efficiently supported \cite{Peng2015Contract}.

To overcome the aforementioned challenges, the concept of heterogeneous C-RANs (H-CRANs) was proposed in \cite{peng2014heterogeneous}.  In H-CRANs, HetNets and C-RANs complement each other. Specifically, Macro BSs are connected to the BBU pool via backhaul with X2/S1 interfaces and RRHs are connected to the BBU pool through wired/wireless fronthaul. In addition to guaranteeing backward compatibility with existing cellular systems and providing ubiquitous connections, BSs are also responsible for delivering control signals and supporting low data-rate services. With the help of macro BSs, unnecessary handover and user re-association can be avoided. On the other hand, RRHs serve high data-rate applications in dedicated zones \cite{tang2015cross}. Besides, the data and control planes are decoupled and the delivery of control signals is shifted from RRHs to macro BSs,  thus the signalling overhead is reduced and the burden on fronthaul links is alleviated.

However, the constrained fronthaul capacity is still a bottleneck in H-CRANs, due to the large number of UEs and the increasing demands for high data-rate service in hotspots. Optical fiber links cannot be used violently because they are expensive.  Wireless fronthaul is thought as an economic choice, but the spectrum resource is scarce. Therefore, a sharing strategy where fronthaul links share the spectrum resource with radio access networks has attracted  lots of attention recently. Radio access network spectrum-based fronthauling has  low sensitivity to propagation conditions, wider coverage, and the reusability of existing equipment.
Many works, e.g., references \cite{Yang2016Energy,zhang2017downlink,Nguyen2016Resource,Wang2016Joint,Xia2016Bandwidth}, have investigated spectrum resource allocation between radio access networks and backhaul (fronthaul) links.  References \cite{Yang2016Energy,zhang2017downlink} studied energy efficiency of HetNets with wireless backhaul, which accounted for the bandwidth and power allocated between macro cells, small cells, and backhaul. Reference \cite{Nguyen2016Resource} considered the joint design of transmit beamforming, power allocation, and spectrum splitting factors that took into account both uplink and downlink transmissions, then formulated a problem of maximizing the achievable sum-rate (SR). A joint problem of cell association and bandwidth allocation for wireless backhaul was optimized in \cite{Wang2016Joint}. Bandwidth allocation combined with interference mitigation techniques was optimized to maximize system SR using the large-dimensional random matrix theory in \cite{Xia2016Bandwidth}.

Besides, compressed-and-forward schemes are another effective way to deal with the challenge of limited fronthaul capacity \cite{park2014performance,park2014fronthaul}. In the compressed-and-forward schemes, signals first are compressed at the BBU pool (or the RRHs) using point-to-point (P2P) compression  or Wyner-Ziv (WZ) coding and then transmitted to the RRHs (or the BBU pool) in downlink (or uplink) via fronthaul. Therefore, the communication rates between the BBU pool and RRHs are reduced. Many works, such as references \cite{zhou2014optimized,zhou2016fronthaul,park2013robust,vu2017adaptive}, have studied fronthaul compression for the uplink C-RAN. In \cite{zhou2014optimized}, a weighted SR was maximized via the optimization of compression noise. In \cite{zhou2016fronthaul}, the authors jointly designed fronthaul compression and beamforming to maximize the achievable SR in the uplink C-RAN with multi-antenna RRHs. An optimization problem was formulated in \cite{park2013robust}, where the compression and BS selection were performed jointly by introducing a sparsity-inducing term into the objective function.  For the downlink C-RAN, the authors in \cite{Park2013Joint,Park2014Inter} studied the joint design of precoding and backhaul compression to maximize the system SR. Besides, a brief overview of fronthaul compression for both uplink and downlink was presented in \cite{simeone2016cloud,park2014fronthaul}.  However, most of these works assumed that the fronthaul was composed of  fiber links with high cost and there was no consideration of spectrum resource sharing between radio access networks and fronthaul links. Moreover, these works were performed according to small-scale fading which varies in the order of milliseconds, thus resulting in  much communication overhead for collecting channel information and high computational complexity to perform optimization in each coherence time of wireless channels.

Motivated by these facts,  we aim to maximize the achievable ergodic SR of uplink H-CRANs with less complexity, overhead, and cost. This goal  is also in line with the expectations of  operators. One possible solution is to  design the spectrum sharing strategy and compression noise simultaneously.  This paper focuses on the uplink transmission of a H-CRAN, where many RRHs are embedded into a macro cell with a multiple-antenna BS. For economic reasons, only the BS are connected to the BBU pool via optical fiber links whereas the RRHs are connected to the BBU pool through wireless fronthaul links. Furthermore, the spectrum resource is shared between the radio access network and fronthaul links. Because the  capacity of fronthaul links is limited, a two-stage compress-and-forward scheme is adopted. At the BBU side, the quantization bits are decompressed, followed by user message decoding.  The contributions of this study are listed as follows.
%In this work, we aim to maximize the achievable  sum-rate (SR) by  jointly optimizing of bandwidth allocation between fronthaul links and access networks and compression noise.

\begin{itemize}
\item In this work, the spectrum sharing strategy and the compress-and-forward scheme are considered together to deal with the challenge of constrained fronthaul capacity.  We formulate a joint optimization problem of bandwidth allocation and compression noise to maximize the achievable ergodic SR, which is a mixed time-scale issue. Because the bandwidth allocation is executed in a large time span whereas the compression noise design is performed in each small time span due to the dependence on small-scale fading. Besides, we consider not only different compress-and-forward schemes, i.e., P2P compression and WZ coding, but also different decoding strategies, i.e., the linear receptions with and without successive interference cancellation (SIC). These different compress-and-forward schemes and decoding strategies have a tradeoff between performance and complexity.
\item In contrast to existing works \cite{zhou2014optimized,zhou2016fronthaul,Park2013Joint,Park2014Inter} where the SR maximization problems were based on fast-changing small-scale fading, we derive the deterministic approximation for the ergodic SR using large-dimensional random matrix theory. Then, an approximate problem of the original problem is introduced, which is  a slow time-scale issue and only depends on statistical channel information. Therefore, the approximate problem can be solved with less communication overhead for obtaining channel information and lower computational complexity.

\item On the basis of deterministic expressions, we first give solutions to the two sub-problems of the joint optimization problem, then propose an algorithm based on Dinkelbach's algorithm to find the optimal solution to the approximation problem under different  compression schemes and decoding strategies. Furthermore, we also propose a low-complexity algorithm to find the near optimal solution to the case of high signal-to-quantization-noise ratio (SQNR).
\end{itemize}

The remainder of this paper is organized as follows. In Section \ref{systemmodel}, we describe the system model and formulate the optimization problem. Then the deterministic approximation of the ergodic SR is given in Section \ref{deterministicequivalents}. Based on the asymptotic result,  we propose the algorithms to find  solutions under different decoding schemes in Sections \ref{analysis1} and \ref{analysis2}, followed by the analysis in special cases in Section \ref{subanalysis}. In Section \ref{numerical results}, simulation results are presented and discussed. Finally, conclusion is drawn in Section \ref{conclusion}.

{\bf Notations}: The notations are given as follows. Matrices and vectors are denoted by bold capital and lowercase symbols. $(\qA)^T$, $(\qA)^H$, and $\tr(\qA)$ stand for transpose, conjugate transpose, and trace of $\qA$, respectively. $\qA\succeq\qzero (\qA\succ\qzero)$ indicates that $\qA$ is a Hermitian positive semidefinite (definite) matrix. For a matrix $\qA=[\qa_1,\ldots,\qa_K]\in\mathbb{C}^{N\times K}$, $\qA_{[m]}=[\qa_1,\ldots,\qa_{m-1},\qa_{m+1},\ldots,\qa_K]\in\mathbb{C}^{N\times K-1}$. The notations $\Ex(\cdot)$ and $\|\cdot\|$ are expectation and Euclidean norm operators, respectively. Finally,
$\qa\sim\mathcal{CN}(\qzero,\bm{\Sigma})$ is a complex Gaussian vector with zero-mean and covariance matrix $\bm{\Sigma}$.

\section{System Model}\label{systemmodel}
\begin{figure}
\includegraphics[width=0.65\textwidth]{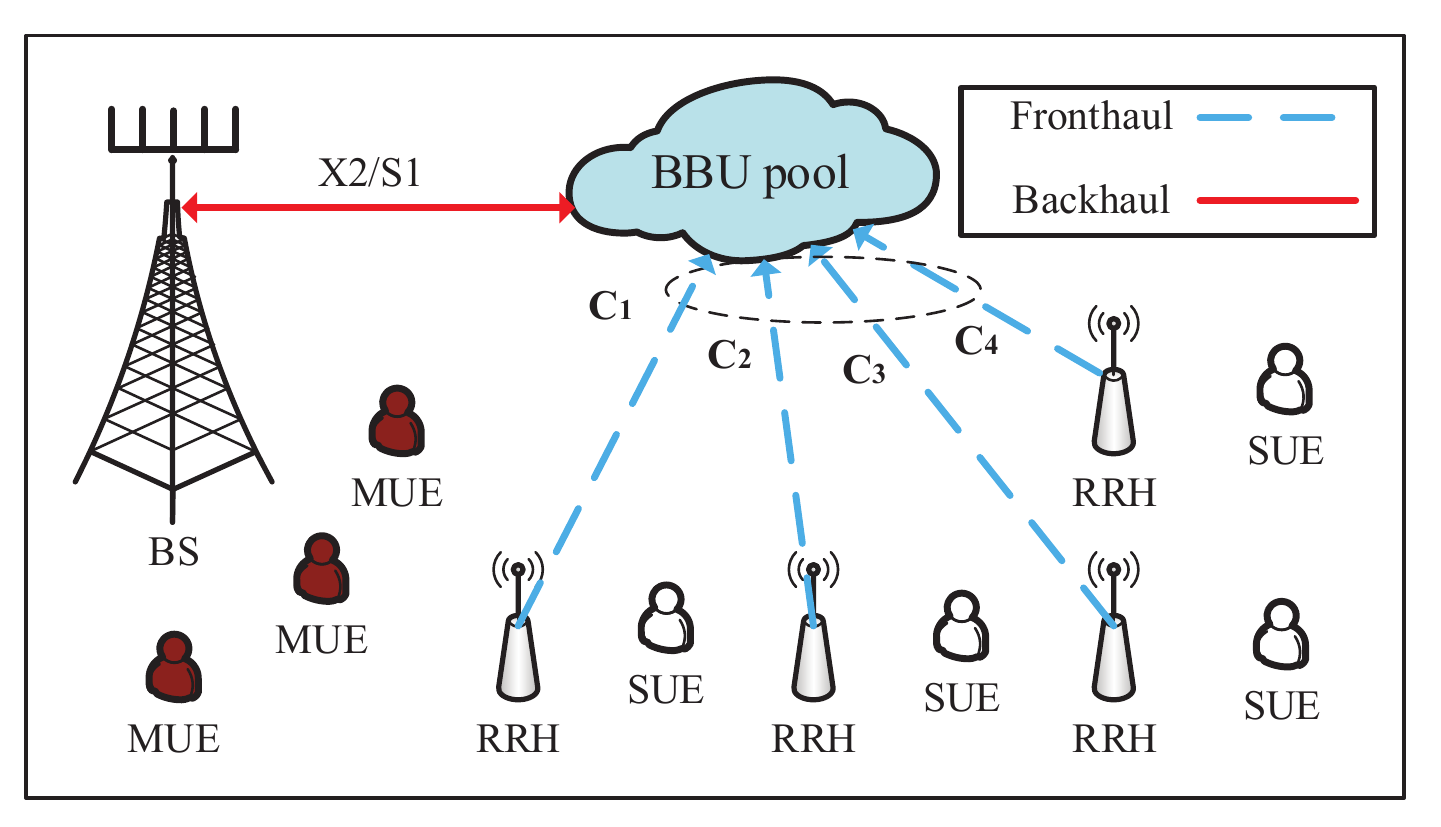}
\centering
\caption{System architecture of a H-CRAN.} \label{system model}
\end{figure}

As shown in Fig. \ref{system model}, an uplink two-tier H-CRAN is considered, consisting of one $N$-antenna macro BS, $L$ single-antenna RRHs, and one BBU pool. The BS is connected to the BBU pool using control and data interfaces, denoted as X2 and S1, via optical fiber links. The RRHs are connected to the BBU pool through noiseless wireless fronthaul links of capacity $C_i$ in bps, $i=1,2,...,L$, which  meet the constraint $\sum_{i=1}^{L}{C_i}\leq C$ \cite{zhou2014optimized}. Such a constraint can model the scenario where the RRHs access to the BBU pool via frequency/time division access scheme, and the number of access slots (frequency or time slots) shared among the RRHs is fixed and limited.
The BS, acting as a centre controller, supports seamless coverage and provides service for $K$ macro UEs (MUEs). The RRHs cooperate with each other and serve as hotspots for $J$ small-cell UEs (SUEs). The MUEs and SUEs each have a single antenna and they simultaneously send messages to BSs and RRHs, respectively. We assume the messages sent from the MUEs can be received by both the BS and RRHs but SUEs' messages cannot be received by the BS directly (e.g., these SUEs are within far areas from the BS or shadow areas of buildings). SUEs' messages are transmitted to the BBU pool via the RRHs for processing and MUEs' messages are processed at the BS and  BBU pool simultaneously. Finally, the processed signals are forwarded to the core network through the BS.   Furthermore, owing to the limited fronthaul capacity, we assume a compress-and-forward scheme is implemented at these RRHs. More specifically, the RRHs first compress the received messages from the UEs and then transmit the quantized description toward the BBU pool. The BBU processor first decompresses the compressed version and then decodes the UEs' messages.
%In this paper, perfect CSI is assumed to be available.

In the frequency domain, the dynamic allocation of bandwidth resource for wireless fronthaul and user communication is
adopted. We assume that all the  available bandwidth for the whole network is $F$ Hz, which can be divided into two orthogonal  parts $F_1=\eta F$ and $F_2=F-F_1$, such that $\eta\in[0,1]$. The former is employed by the BS and RRHs to serve the MUEs and SUEs and the latter is dedicated to fronthaul links. It's assumed that $C=C_0F_2$ where $C_0$ in bps/Hz is constant and depends on the specific propagation environment. Without losing generality, we normalize the fronthaul capacity constraint as $\sum_{i=1}^L{\bar{C}_i}\leq (1-\eta)C_0$ where $\bar{C}_i=C_i/F$ in bps/Hz.
\subsection{Achievable SR at BS }
As discussed above, we assume the messages transmitted from the SUEs are only received at the RRHs while ones from the MUEs can be detected at both the BS and RRHs. Since we are concerned about the achievable SR of the MUEs and SUEs, we simply divide the achievable SR into two parts: one at the BS and the other at the central BBU pool. We first focus on the achievable SR at the BS.

The received signal $\qy^{(B)}\in\bbC^{N\times 1}$ at the BS from the MUEs through sub-frequency band $F_1$ is expressed as
\begin{equation}
  \qy^{(B)}=\qH\qx^{(M)}+\qn^{(B)},
\end{equation}
where $\qH=[\qh_1,\ldots,\qh_K]\in\bbC^{N\times K}$ is the channel matrix between the MUEs and BS,  $\qh_k=\sqrt{N\nu_k}\tilde{\qh}_k$, $\tilde{\qh}_k\sim\calCN(\qzero, \frac{1}{N}\qI_N)$ and $\nu_k$  represent the small-scale and large-scale fading coefficients, respectively \cite{chi2017message}, $\qn^{(B)}\sim\calCN(\qzero,\sigma^2\qI_N)$ describes the independent additive white Gaussian noise (AWGN) \cite{Liu2016Gaussian}, $\qx^{(M)}=[x^{(M)}_1,\ldots,x^{(M)}_K]^T\in\bbC^{K\times 1}$ is the signal vector of the MUEs with $x^{(M)}_k=\sqrt{p^{(M)}_k}s^{(M)}_k$, $s^{(M)}_k\sim\calCN(0,1)$ denotes the transmitted symbols of MUE $k$, and $p^{(M)}_k$ is the transmit power of MUE $k$.
%$p^{(M)}_k=\tilde{p}^{(M)}_k/\eta$ with $\tilde{p}^{(M)}_k$ being the transmit power of MUE $k$

Under the linear reception, the achievable ergodic SR of the MUEs at the BS in bps/Hz is computed as
\begin{equation}
R^{(B,L)}(\eta)=\eta R^{(B,L)}_0=\eta\sum\limits_{k=1}^K{R^{(B,L)}_{0,k}},
\end{equation}
where
\begin{equation}
  R^{(B,L)}_{0,k}=\Ex\left(\log_2\frac{|\qH\qP^{(M)}\qH^H+\sigma^2\qI_N|}{|\qH_{[k]}\qP^{(M)}_{[k]}\qH^H_{[k]}+\sigma^2\qI_N|}\right),
\end{equation}
in which $\qP^{(M)}=\text{diag}(\qp^{(M)})$ with $\qp^{(M)}=[p^{(M)}_1,\ldots,p^{(M)}_K]^T$. Note that the symbol ``L'' in the superscript suggests the linear reception.
However, if the SIC method is applied, without losing generality, the decoding order is assumed as $1,2,\ldots,K$. Then the achievable SR at the BS becomes \cite{simeone2016cloud}
\begin{align}
R^{(B,SIC)}(\eta)=\eta R^{(B,SIC)}_0&=\eta\Ex\left(\sum\limits_{k=1}^K{\log_2\frac{|\sum_{k^{\prime}=k}^K{p^{(M)}_{k^{\prime}}\qh_{k^{\prime}}\qh_{k^{\prime}}^H}+\sigma^2\qI_N|}{|\sum_{k^{\prime}=k+1}^K{p^{(M)}_{k^{\prime}}\qh_{k^{\prime}}\qh_{k^{\prime}}^H}+\sigma^2\qI_N|}}\right),\\
&=\eta\Ex\left(\log_2\frac{|\qH\qP^{(M)}\qH^H+\sigma^2\qI_N|}{|\sigma^2\qI_N|}\right).
\end{align}
Obviously, the achievable SR under the linear reception with SIC is greater than that under the linear reception without SIC due to less interference. But this advantage comes at the cost of higher computational complexity introduced by the SIC operation.

\subsection{Achievable SR at BBU Pool}\label{subsection_3}
The received signals at the BBU pool includes those from both the SUEs and MUEs. To the RRHs, the MUEs and SUEs are equivalent. Therefore, for ease of expression, we define $\qG=[\qG^{(S)}, \qG^{(M)}]\in\bbC^{L\times (J+K)}$ as a composite channel matrix, where $\qG^{(S)}$ and $\qG^{(M)}$ represent the channel matrix between the SUEs and RRHs and that between the MUEs and RRHs, respectively. $\qG^{(S)}=[\qg^{(S)}_1,\ldots,\qg^{(S)}_J]$, where $\qg^{(S)}_j=[g^{(S)}_{j1},\ldots,g^{(S)}_{jL}]^{T}$, $g^{(S)}_{jl}=\sqrt{L\mu_{jl}^{(S)}}\tilde{g}^{(S)}_{jl}$ represents the channel coefficient between the $j$-th SUE and the $l$-th RRH, and  $\mu_{jl}^{(S)}$ and $\tilde{g}^{(S)}_{jl}\sim\calCN(0,\frac{1}{L})$ describe the corresponding large-scale and small-scale fading coefficients, respectively. $\qG^{(M)}$ is similar to the definition of $\qG^{(S)}$. We further define $\qx=[(\qx^{(S)})^T ,(\qx^{(M)})^T]^T$  as a composite signal vector, where $\qx^{(S)}$ and $\qx^{(M)}$ are the signal vectors from the SUEs and  MUEs, respectively. $\qx^{(S)}=[x^{(S)}_1, \ldots, x^{(S)}_J ]^T$, where $x^{(S)}_j=\sqrt{p^{(S)}_j} s^{(S)}_j$, $s^{(S)}_j\sim\calCN(0,1)$ is the signal transmitted by SUE $j$, and $p^{(S)}_j$ denotes the  transmit power of SUE $j$.  Thus the received signal at RRH $l$ is expressed as
%$p^{(S)}_j=\tilde{p}^{(S)}_j/\eta$ with $\tilde{p}^{(S)}_j$ denoting the  transmit power of SUE $j$
\begin{equation}
  y^{(R)}_l=\sum\limits_{m=1}^{M}{g_{lm}x_m}+n^{(R)}_l,
\end{equation}
where $M=J+K$, $n^{(R)}_l\sim\calCN(0,\sigma^2)$ represents the AWGN at AP $l$, $x_m$ denotes the $m$-th element of $\qx$,  and $g_{lm}$ is the $(l,m)$-th element  of $\qG$.

We assume the compress-and-forward scheme  is applied at the RRHs, e.g., point-to-point (P2P) compression \cite{zhou2013approximate} or Wyner-Ziv (WZ) coding \cite{zhou2014optimized}. The RRHs first quantizes the received signal $\qy^{(R)}=[y^{(R)}_1,\ldots,y^{(R)}_L]^T$ into $\check{\qy}^{(\kappa)}=[\check{y}^{(\kappa)}_1,\ldots,\check{y}^{(\kappa)}_L]^T$ (the superscript ``$\kappa$'' can be P2P or WZ indicating different quantization schemes adopted), and then transmits the compressed version to the BBU pool for central processing through  wireless fronthaul links. At the BBU side, the quantization codewords are decompressed and then the messages are decoded \cite{zhou2016fronthaul}.
In this work, we obtain the achievable rate region in the case where each RRH only compresses its own received signal. Similar to reference \cite{park2014fronthaul}, we model the relationship between the received signal $y^{(R)}_l$ and its compressed description $\check{y}^{(\kappa)}_l$  as Gaussian test channel,
\begin{equation}
  \check{y}^{(\kappa)}_l=y^{(R)}_l+q_l,
\end{equation}
where $q_l\sim\calCN(0,\psi_l^2)$ is independent quantization noise and $\psi_l^2$ describes the variance of the quantization noise at RRH $l$.

In the case of the linear reception  at the BBU pool, the achievable ergodic SR is
\begin{equation}\label{achievable rate at bbu with lmmse}
R^{(P,L)}(\eta,\bm{\Psi})=\eta R^{(P,L)}_0(\bm{\Psi})=\eta\sum\limits_{m=1}^M{R^{(P,L)}_{0,m}}(\bm{\Psi}),
\end{equation}
where
\begin{equation}
  R^{(P,L)}_{0,m}(\bm{\Psi})=\Ex\left(\log_2\frac{|\qG\qP\qG^H+\bm{\Psi}+\sigma^2\qI_L|}{|\qG_{[m]}\qP_{[m]}\qG_{[m]}^H+\bm{\Psi}+\sigma^2\qI_L|}\right),
\end{equation}
in which $\bm{\Psi}$ is an $L\times L$ diagonal matrix with diagonal elements $\{\psi_l^2\}$'s, $\qP=\text{diag}(\qp)$ with $\qp=[(\qp^{(S)})^T, (\qp^{(M)})^T]^T$,  and $\qp^{(S)}=[p^{(S)}_1,\ldots,p^{(S)}_J]^T$.

When the SIC method is employed, without loss of generality, the decoding order is assumed as $1,2,\ldots,M$ \cite{zhang2014the,zhang2017energy,lin2017anew}.  Then the achievable ergodic SR of the SUEs and MUEs at the BBU is expressed as
\begin{align}
  R^{(P,SIC)}(\eta,\bm{\Psi})= \eta R^{(P,SIC)}_0(\bm{\Psi})&=\eta\Ex\left(\sum\limits_{m=1}^M{\log_2\frac{|\sum_{m^{\prime}=m}^M{p_{m^{\prime}}\qg_{m^{\prime}}\qg_{m^{\prime}}^H}+\bm{\Psi}+\sigma^2\qI_L|}{|\sum_{m^{\prime}=m+1}^M{p_{m^{\prime}}\qg_{m^{\prime}}\qg_{m^{\prime}}^H}+\bm{\Psi}+\sigma^2\qI_L|}}\right),\\
  &=\eta\Ex\left(\log_2\frac{|\qG\qP\qG^H+\bm{\Psi}+\sigma^2\qI_L|}{|\bm{\Psi}+\sigma^2\qI_L|}\right)\label{achievable rate at bbu}.
\end{align}

In what follows we give the fronthaul constraints that must be satisfied to achieve the SRs in \eqref{achievable rate at bbu} and \eqref{achievable rate at bbu with lmmse} under two different compression strategies, i.e., P2P compression and WZ coding, respectively.

\subsubsection{P2P Compression Scheme}
As a consequence of P2P compression, each RRH produces a binary string that allows the decompressor unit at the BS to identify the quantized signal from a certain codebook in parallel \cite{park2014fronthaul}.  According to rate-distortion theory, the signal $\check{y}^{(P2P)}$ can be recovered if the fronthaul rate satisfies the condition
\begin{align}
  R^{(fh,P2P)}(\eta,\bm{\Psi})&=\eta R^{(fh,P2P)}_0(\bm{\Psi})
   =\eta\Ex\left(\sum\limits_{l=1}^L{\log_2\frac{\sum_{m=1}^{M}{p_m|g_ {lm}|^2}+\sigma^2+\psi_l^2}{\psi_l^2}}\right)
  %&=\eta\Ex\log_2\frac{|\diag(\qG\qP\qG^H)+\bm{\Psi}+\sigma^2\qI_L|}{|\bm{\Psi}|}\\
  \leq (1-\eta)C_0. \label{p2p constraint}
\end{align}
%where $\bm{\Psi}=\diag(\psi_1,\psi_2,\ldots,\psi_L)$ is a diagonal matrix since the compression and decompression are carried out in parallel across different RRHs under P2P compression scheme.
This constraint suggests that, from an intuitive level, larger $\psi_l^2$'s cause a smaller $R^{(fh,P2P)}_0(\eta,\bm{\Psi})$ and thus less bandwidth is required by fronthaul.
However, such separate and independent processing does not take into account the statistical correlation across the signals $y^{(R)}_l$, $l=1,\ldots,L$, received at different RRHs \cite{simeone2016cloud}, which can be used as side information when the BBU decompresses the quantized signals from the RRHs. WZ coding provides an efficient way to leverage the side information available at the RRHs.

\subsubsection{WZ Coding Scheme}
Taking advantage of the correlation of the received signals at all the RRHs which results from the mutual interference between UEs, WZ coding can achieve a higher performance and make better use of limited fronthaul capacity than P2P compression  \cite{park2014fronthaul}. WZ coding enables the compressor to use a finer quantizer and associate the same binary string to a subset of codewords, whereas P2P compression associates a distinct
binary string  with each codeword in the quantization codebook. According to \textbf{Proposition 1} in \cite{zhou2014optimized},  the SRs in \eqref{achievable rate at bbu} and \eqref{achievable rate at bbu with lmmse} can be achieved if the condition
\begin{equation}
  R^{(fh,WZ)}(\eta,\bm{\Psi})=\eta R^{(fh,WZ)}_0(\bm{\Psi})=\eta\Ex\left(\log_2\frac{|\qG\qP\qG^H+\bm{\Psi}+\sigma^2\qI_L|}{|\bm{\Psi}|}\right)\leq (1-\eta)C_0, \label{wyner constraint}
\end{equation}
is satisfied.
\subsection{Problem Formulation}
By jointly designing the compression noise variance matrix $\bm{\Psi}$ and the bandwidth allocation factor $\eta$, the achievable ergodic SR maximization problem can be formulated as:
\begin{subequations}\label{orginal problem}
\begin{align}
   \mathop{\max}_{\eta,\bm{\Psi}}\quad  &R^{(B,\varsigma)}(\eta,\bm{\Psi})+ R^{(P,\varsigma)}(\eta,\bm{\Psi})  \label{optimization problem}\\
   \text{s.t.} \quad &R^{(fh,\kappa)}(\eta,\bm{\Psi})\leq(1-\eta)C_0 \label{cond1}, \\
   &\eta\in [0,1]\label{cond2},\\
   &\bm{\Psi}_{ll}\geq 0, \quad l=1,2,\ldots L, \label{cond3}\\
   &\bm{\Psi}_{l^{\prime}l}=0, \quad l^{\prime}\neq l,   \label{cond4}
\end{align}
\end{subequations}
where the symbol ``$\varsigma$" in the superscript can be ``L'' or ``SIC'' representing different decoding strategies (i.e., the linear receptions with and without SIC), the fronthaul constraint \eqref{cond1}  can be either \eqref{p2p constraint} or \eqref{wyner constraint}, depending on the specific compression strategy. Constraints \eqref{cond3} and \eqref{cond4} indicate that $\bm{\Psi}$ is a diagonal matrix. The objective function is a difference-of-convex problem and some algorithms  are proposed in \cite{beck2009gradient,zhou2014optimized,Park2013Joint} to solve related problems. However, since the optimization of ($\eta$, $\bm{\Psi}$) is based on the ergodic SR, Monte Carlo averaging over channels needs  a lot of samples to capture the variations of both large-scale fading and small-scale fading. Because small-scale fading varies at the level of milliseconds, collecting channel information leads to too much communication overhead and calculating average results over channels also brings more prohibitively computational complexity. Besides, the bandwidth allocation problem is a slow time-scale issue because it  is usually executed in a large time span whereas the compression noise design problem  is a fast time-scale issue since it is performed in each small time span (for example, a coherence time interval of wireless channels) due to its dependence on fast-changing small-scale fading. Therefore, the joint problem is a mixed time-scale issue. To address these challenges, we first introduce the asymptotic expression of the ergodic SR in the large-system regime and then find solutions based on the deterministic approximation in the following sections.

\section{Deterministic Equivalents}\label{deterministicequivalents}
In the following, we understand the ergodic SR in the large-system regime \cite{zhang2013large,xia2017large} where $K$, $J$, $N$, and $L$ grow infinitely while keeping fixed ratios $\chi_1 =K / N$ and $\chi_2 =M /L$ such that $0<\liminf_N \chi_1<\limsup_N \chi_1<1$ and $0<\liminf_L\chi_2<\limsup_L \chi_2<1$ \cite{Xia2016Bandwidth}.
For notational convenience, we use $N\rightarrow\infty$ and $L\rightarrow\infty$ to refer to $N,K\rightarrow\infty$ and $L, M\rightarrow \infty$, respectively.
We will derive deterministic approximations $\bar{R}^{(B,\varsigma)}(\eta)$, $\bar{R}^{(P,\varsigma)}(\eta,\bm{\Psi})$, and $\bar{R}^{(fh,\kappa)}(\eta,\bm{\Psi})$ of the ergodic SRs $R^{(B,\varsigma)}(\eta)$, $R^{(P,\varsigma)}(\eta,\bm{\Psi})$, and $R^{(fh,\kappa)}(\eta,\bm{\Psi})$, respectively.
\subsection{Deterministic Equivalents for $R^{(B,SIC)}(\eta)$ and $R^{(B,L)}(\eta)$}\label{subsection1}

Since $\tilde{\qh}_k$'s are independently and identically distributed (i.i.d.) complex Gaussian variables whose real and imaginary parts are independent. According to references \cite{wen2013adetermini,zhang2013oncap},  we have the following lemma.
\begin{lemma}\label{lemma1}
Given that $\tilde{\qh}_k$'s are i.i.d. complex Gaussian variables with independent real and imaginary parts, as $N\rightarrow\infty$, we have $R^{(B,SIC)}(\eta)-\bar{R}^{(B,SIC)}(\eta)\rightarrow0$, where $\bar{R}^{(B,SIC)}(\eta)=\eta\bar{R}^{(B,SIC)}_0$ and
\begin{align}
 \bar{R}^{(B,SIC)}_0%&=\frac{1}{\log2}\left[\log|\bm{\Gamma}|+\sum\limits_{k=1}^K{(\frac{1}{1+e_k}-\log\frac{1}{1+e_k})}-N\log\sigma^2-K\right],\\
 &=\frac{1}{\log2}\left(\Delta_0-\log|\sigma^2\qI_N|\right),\label{asym rp}
\end{align}
where
\begin{equation}
 \Delta_0=\log|\bm{\Gamma}|+\sum\limits_{k=1}^K{(\frac{1}{1+e_k}-\log\frac{1}{1+e_k})}-K,
\end{equation}
in which $e_k=p^{(M)}_{k}\nu_k\tr\left(\bm{\Gamma}^{-1}\right)$ and
%\begin{align}
%  e_k%&=\frac{1}{N}\tr\left((p^{(M)}_k\nu_kN\qI_N)(\sum\limits_{k^{\prime}=1}^{K}{\frac{p^{(M)}_{k^{\prime}}\nu_{k^{\prime}}\qI_N}{1+e_{k^{\prime}}}}+\sigma^2\qI_N)^{-1}\right),\\
%  &=p^{(M)}_{k}\nu_k\tr\bm{\Gamma}^{-1},
%\end{align}

\begin{equation}
 \bm{\Gamma}=\left(\sum\limits_{k=1}^{K}{\frac{p^{(M)}_{k}\nu_k}{1+e_{k}}}+\sigma^2\right)\qI_N.
\end{equation}

In the case of the linear reception without SIC, as $N\rightarrow\infty$, the deterministic equivalent of $R^{(B,L)}$ is given as $R^{(B,L)}(\eta)-\bar{R}^{(B,L)}(\eta)\rightarrow0$, where $\bar{R}^{(B,L)}(\eta)=\eta\bar{R}^{(B,L)}_0=\eta\sum\limits_{k=1}^K{\bar{R}^{(B,L)}_{0,k}}$ and $\bar{R}^{(B,L)}_{0,k}=\frac{1}{\log2}(\Delta_0-\Delta_k)$ in which
\begin{equation}
  \Delta_k= \log|\bm{\tilde{\Gamma}}_k|+\sum\limits_{k^{\prime}=1,k^{\prime}\neq k}^K{(\frac{1}{1+\tilde{e}_{kk^{\prime}}}-\log\frac{1}{1+\tilde{e}_{kk^{\prime}}})}-(K-1),
\end{equation}
with $\tilde{e}_{kk^{\prime}}=p^{(M)}_{k^{\prime}}\nu_{k^{\prime}}\tr\bm{\tilde{\Gamma}}_{k}^{-1}$ and
\begin{equation}
  \bm{\tilde{\Gamma}}_k=\left(\sum\limits_{k^{\prime}=1,k^{\prime}\neq k}^{K}{\frac{p^{(M)}_{k^{\prime}}\nu_{k^{\prime}}}{1+\tilde{e}_{kk^{\prime}}}}+\sigma^2\right)\qI_N.
\end{equation}

\end{lemma}
\IEEEproof  Refer to Appendix \ref{appendixA}.

\subsection{Deterministic Equivalents for $R^{(P,SIC)}(\eta,\bm{\Psi})$ and $R^{(P,L)}(\eta,\bm{\Psi})$}
Based on the analysis of $\bar{R}^{(B,\varsigma)}(\eta)$ in Section \ref{subsection1}, we  use large-dimensional random matrix theory again to derive the deterministic equivalents of $R^{(P,SIC)}(\eta,\bm{\Psi})$ and $R^{(P,L)}(\eta,\bm{\Psi})$  in the following lemma.
\begin{lemma}\label{lemma2}
Given that $\tilde{\qg}_m$'s  are i.i.d. complex Gaussian variables with independent real and imaginary parts, as $L\rightarrow\infty$, we have $R^{(P,SIC)}(\eta,\bm{\Psi})-\bar{R}^{(P,SIC)}(\eta,\bm{\Psi})\rightarrow0$, where $\bar{R}^{(P,SIC)}(\eta,\bm{\Psi})=\eta\bar{R}^{(P,SIC)}_0(\bm{\Psi})$ and
\begin{equation}
 \bar{R}^{(P,SIC)}_0(\bm{\Psi})=\frac{1}{\log2}\left(\nabla_0-\log|\bm{\Psi}+\sigma^2\qI_L|\right),
\end{equation}
where
\begin{equation}\label{nabla}
\nabla_0=\log|\bm{\Lambda}|+\sum\limits_{m=1}^M{(\frac{1}{1+b_m}-\log\frac{1}{1+b_m})}-M,
\end{equation}
in which $b_m=p_m\tr\left(\qU_m\bm{\Lambda}^{-1}\right)$ and
%\begin{align}
%  b_m
%  %&=\frac{1}{N}\tr Np_m\qU_m\left(\frac{1}{N}\sum\limits_{m^{\prime}=1}^M{\frac{Np_{m^{\prime}}\qU_{m^{\prime}}}{1+b_{m^{\prime}}}}+\bm{\Psi}+\omega\qI_L\right)^{-1}\\
%  &=p_m\tr\qU_m\bm{\Lambda}^{-1},
%\end{align}
\begin{equation}\label{Lambda}
  \bm{\Lambda}=\sum\limits_{m=1}^M{\frac{p_{m}\qU_m}{1+b_m}+\bm{\Psi}+\sigma^2\qI_L},
\end{equation}
with $\qU_m=\text{diag}(\qu_m)$ and $\qu_m=[\mu_{1m}, \ldots, \mu_{Lm}]^T$.

In the case of the linear reception without SIC, we have $R^{(P,L)}(\eta,\bm{\Psi})-\bar{R}^{(P,L)}(\eta,\bm{\Psi})\rightarrow0$, where $\bar{R}^{(P,L)}(\eta,\bm{\Psi})=\eta\bar{R}^{(P,L)}_0(\bm{\Psi})=\eta\sum\limits_{m=1}^M{\bar{R}^{(P,L)}_{0,m}}(\bm{\Psi})$ and $\bar{R}^{(P,L)}_{0,m}(\bm{\Psi})=\frac{1}{\log2}(\nabla_0-\nabla_m)$ in which
\begin{equation}
  \nabla_m=\log|\bm{\tilde{\Lambda}}_m|+\sum\limits_{m^{\prime}=1,m^{\prime}\neq m}^M{(\frac{1}{1+b_{mm^{\prime}}}-\log\frac{1}{1+b_{mm^{\prime}}})}-(M-1),
\end{equation}
with $\tilde{b}_{mm^{\prime}}=p_{m^{\prime}}\tr\left(\qU_{m^{\prime}}\bm{\tilde{\Lambda}}_m^{-1}\right)$ and
\begin{equation}
  \bm{\tilde{\Lambda}}_m=\sum\limits_{m^{\prime}=1,m^{\prime}\neq m}^M{\frac{p_{m^{\prime}}\qU_{m^{\prime}}}{1+\tilde{b}_{mm^{\prime}}}+\bm{\Psi}+\sigma^2\qI_L}.
\end{equation}
\end{lemma}

\IEEEproof Refer to the proof of $\bar{R}^{(B,\varsigma)}(\eta)$ in Appendix \ref{appendixA}, due to the similarity between $R^{(B,\varsigma)}(\eta)$ and $R^{(P,\varsigma)}(\eta,\bm{\Psi})$.

\subsection{Deterministic Equivalents for $R^{(fh,WZ)}(\eta,\bm{\Psi})$ and $R^{(fh,P2P)}(\eta,\bm{\Psi})$}
Then, we also give two deterministic equivalents of fronthaul constraints $R^{(fh,P2P)}(\eta,\bm{\Psi})$ and $R^{(fh,WZ)}(\eta,\bm{\Psi})$ for two different compression strategies, i.e., P2P compression and WZ coding.
\subsubsection{P2P compression}
%We first rewrite $\tr(\diag(\qG\qP\qG^H)+\sigma^2\qI_L)^{-1}$ as
%\begin{equation}
%  \diag(\qG\qP\qG^H)=\qW\qW^H
%\end{equation}
%where $\qW=[\qW^1,\ldots,\qW^M]\in\bbC^{L\times ML}$ is a collection of matrices $\{\qW^m\}$'s,  which are defined as
%\begin{equation}
%  \qW^m=\sqrt{p_m}\diag(g_{lm})_{l=1}^L
%\end{equation}
%whose $l$-th column is
%\begin{equation}
%  \qw^m_l=[0,0,\ldots,\sqrt{p_m}g_{lm},\ldots,0]^T=\sqrt{N\bm{\vartheta}^m_l}\tilde{\qg}_m
%\end{equation}
%where $\bm{\vartheta}^m_l=p_m\mu_{ml}\qI_l$.
%Then based on the proof of $R^{(B)}$, the deterministic approximation of $R^{(bh)}$ in \eqref{p2p constraint} is computed by
%\begin{equation}
%  R^{(bh)}(\sigma^2)-\bar{R}^{(bh)}(\sigma^2)\rightarrow0
%\end{equation}
%where
%\begin{equation}
% \bar{R}^{(bh)}(\sigma^2)=\frac{\eta}{\log2}\left[\log|\bm{\Pi}|+\sum\limits_{m=1}^M{(\log\frac{1}{1+c_m}-\frac{1}{1+c_m})}-\log|\bm{\Psi}|\right]
%\end{equation}
%where
%\begin{equation}
%  \bm{\Pi}=\frac{1}{L}\sum\limits_{k=1}^{ML}{\frac{\bm{\Theta}^{f_1(k)}_{f_2(k)}}{1+c_k}+\bm{\Psi}+\sigma^2\qI_L}
%\end{equation}
%and
%\begin{equation}
%  c_m=\tr\bm{\Theta}^{f_1(m)}_{f_2(m)}\bm{\Pi}^{-1}
%\end{equation}
%where $\bm{\Theta}^m_l=\mu_{lm}\diag(\qI_l)$ $f_1(k)$ and $f_2(k)$ are functions of $k$ and given as $f_1(k)=\lfloor \frac{k}{L}\rfloor$ and
%\begin{equation}
%  f_2(k)=
%  \begin{cases}
%     k \bmod L, &\text{if $(k \bmod L)\neq 0$,}\\
%     L, &\text{else.}
%  \end{cases}
%\end{equation}
Observe that if $M$ is large enough, according to  the law of large numbers, $\sum_{m=1}^{M}{p_m|g_ {lm}|^2}$ converges to its mean,
\begin{equation}
  \frac{1}{M}\sum\limits_{m=1}^{M}{p_m|g_ {lm}|^2}\xrightarrow{M\rightarrow\infty}\frac{1}{M}\sum\limits_{m=1}^{M}{p_m\mu_ {lm}},\label{law of large number}
\end{equation}
where $\mu_ {lm}$ is the large-scale fading coefficient between RRH $l$ and UE $m$.  Therefore, the deterministic approximation of $R^{(fh,P2P)}(\eta,\bm{\Psi})$ in \eqref{p2p constraint} is given by the following lemma.
\begin{lemma}
According to the law of large numbers, we have the result in \eqref{law of large number} and $R^{(fh,P2P)}(\eta,\bm{\Psi})-\bar{R}^{(fh,P2P)}(\eta,\bm{\Psi})\xrightarrow{M\rightarrow \infty}0$, where $\bar{R}^{(fh,P2P)}(\eta,\bm{\Psi})=\eta\bar{R}^{(fh,P2P)}_0(\bm{\Psi})$ and
\begin{align}
   \bar{R}^{(fh,P2P)}_0(\bm{\Psi})&=\sum\limits_{l=1}^L{\log_2\frac{\sum\limits_{m=1}^{M}{p_m \mu_ {lm}}+\sigma^2+\psi_l^2}{\psi_l^2}}.
   %&=\eta\log_2\frac{|\sum\limits_{m=1}^M{p_m\qU_m}+\bm{\Psi}+\sigma^2\qI_L|}{|\bm{\Psi}+\sigma^2\qI_L|}.
\end{align}
\end{lemma}

\subsubsection{WZ compression}
Due to the similarity of $R^{(P,SIC)}(\eta,\bm{\Psi})$ and $R^{(fh,WZ)}(\eta,\bm{\Psi})$, we directly give the deterministic equivalent of $R^{(fh,WZ)}(\eta,\bm{\Psi})$  as the following lemma according to \textbf{Lemma} \ref{lemma2}.
\begin{lemma}
Given that $\tilde{\qg}_m$'s  are i.i.d. complex Gaussian variables with independent real and imaginary parts, as $L\rightarrow\infty$, we have $R^{(fh,WZ)}(\eta,\bm{\Psi})-\bar{R}^{(fh,WZ)}(\eta,\bm{\Psi})\rightarrow0$,
where $\bar{R}^{(fh,WZ)}(\eta,\bm{\Psi})=\eta\bar{R}^{(fh,WZ)}_0(\bm{\Psi})$ and
\begin{equation}
 \bar{R}^{(fh,WZ)}_0(\bm{\Psi})=\frac{1}{\log2}\left(\nabla_0-\log|\bm{\Psi}|\right),
\end{equation}
in which $\nabla_0$ is given in \eqref{nabla}.
\end{lemma}
Note that one thing these approximations have in common is that they are determined by statistical channel knowledge (i.e., large-scale fading), instead of small-scale fading. Generally, statistical channel knowledge varies much slower than small-scale fading. Thus the communication overhead for obtaining channel information is reduced significantly.

\section{Joint Design under Linear Reception with SIC}\label{analysis1}
Problem \eqref{orginal problem} has different forms because different decoding strategies and  compression schemes are included simultaneously. For analytical ease, problem \eqref{orginal problem} is separated into two problems according to decoding strategies  and we first consider the case of the linear reception with SIC. By replacing $R^{(B,SIC)}(\eta)$, $R^{(P,SIC)}(\eta,\bm{\Psi})$, and $R^{(fh,\kappa)}(\eta,\bm{\Psi})$ with their deterministic equivalents $\bar{R}^{(B,SIC)}(\eta)$, $\bar{R}^{(P,SIC)}(\eta,\bm{\Psi})$, and $\bar{R}^{(fh,\kappa)}(\eta,\bm{\Psi})$, respectively,  we introduce an alternative to  problem \eqref{orginal problem} as follows:
\begin{subequations}\label{sym objec problem}
\begin{align}
  \mathop{\max}_{\eta,\bm{\Psi}} \quad &  \bar{R}^{(B,SIC)}(\eta,\bm{\Psi})+ \bar{R}^{(P,SIC)}(\eta,\bm{\Psi})\label{asym objec function}\\
   \text{s.t.}\quad & \bar{R}^{(fh,\kappa)}(\eta,\bm{\Psi})\leq(1-\eta)C_0, \label{cond5} \\
   &\eqref{cond2}-\eqref{cond4},\label{cond6}
\end{align}
\end{subequations}
which avoids redundant computation resulting from Monte Carlo averaging over channels, because the objective function and fronthaul constraint are only dependent on statistical channel knowledge and irrelevant to small-scale fading. Furthermore, the joint optimization problem \eqref{orginal problem}  of the bandwidth allocation and compression noise design now turns into a slow time-scale issue and we can  find solutions to the joint optimization problem \eqref{sym objec problem} in a large time span, not each small time span. However, finding the global optimal results of problem \eqref{sym objec problem} is still a challenge due to the non-convexity of the objective function over variable matrix $\bm{\Psi}$. In what follows, we first analyze two sub-problems of problem \eqref{sym objec problem}: 1) find the optimal $\bm{\Psi}$ given that $\eta$ is fixed, and 2) find the optimal $\eta$ given that $\bm{\Psi}$ is fixed. Then we reformulate problem \eqref{sym objec problem} as a concave-convex fractional programming \cite{dinkelbach1967nonlinear,zappone2015energy}. Finally, an algorithm is proposed to find solutions to the joint optimization problem of $\bm{\Psi}$ and $\eta$.

\subsection{Optimization of Fronthaul Compression}
Given a fixed $\eta$ value, We find the optimal value of $\bm{\Psi}$ by solving  the following sub-problem:
\begin{subequations}\label{subpro11}
\begin{align}
  \mathop{\max}_{\bm{\Psi}} \quad &  \bar{R}^{(P,SIC)}(\bm{\Psi}) \label{subpro1}\\
   \text{s.t.}\quad & \eqref{cond5},\ \eqref{cond3}, \ \text{and} \ \eqref{cond4},
\end{align}
\end{subequations}
which is still non-convex because the objective function $\bar{R}^{(P,SIC)}(\bm{\Psi}) $  is actually a  convex function, instead of a concave function, with respect to $\bm{\Psi}$.  Thanks to the following lemma, which sheds light on solving problem \eqref{subpro11}.
\begin{lemma}[\cite{li2013transmit}]
 Given an $L\times L$ complex matrix $\bm{\Gamma}\succ\qzero$. Consider the function $f(\bm{\Omega})=-\tr(\bm{\Omega}\bm{\Gamma})+\log|\bm{\Omega}|+L$, then
\begin{equation}
  \log|\bm{\Gamma}^{-1}|=\mathop{\max}_{\bm{\Omega}\in\bbC^{L\times L},\bm{\Omega}\succeq\qzero}f(\bm{\Omega}),
\end{equation}
 with the optimal solution $\bm{\Omega}^{\star}=\bm{\Gamma}^{-1}$.
\end{lemma}
From \textbf{Lemma 5}, the inequality
\begin{equation}\label{inequality}
  -\log|\bm{\Gamma}|\geq\log|\bm{\Omega}|-\tr(\bm{\Omega}\bm{\Gamma})+L,
\end{equation}
always holds and the equality holds with $\bm{\Omega}=\bm{\Gamma}^{-1}$.
By applying inequality \eqref{inequality} to the second term of $\bar{R}^{(P,SIC)}(\bm{\Psi})$ via setting $\bm{\Gamma}=\bm{\Psi}+\sigma^2\qI_L$, then we reformulate  sub-problem \eqref{subpro11} as
 \begin{subequations}\label{appr_subpro1 problem}
\begin{align}
 \mathop{\max}_{\bm{\Psi},\bm{\Omega}} \quad  &\tilde{\bar{R}}^{(P,SIC)}(\bm{\Psi},\bm{\Omega})\label{appr_subpro1}\\
 \text{s.t.}\quad & \eqref{cond5},\ \eqref{cond3}, \ \text{and} \ \eqref{cond4},
\end{align}
\end{subequations}
where $\tilde{\bar{R}}^{(P,SIC)}(\bm{\Psi},\bm{\Omega})=\eta\tilde{\bar{R}}^{(P,SIC)}_0(\bm{\Psi},\bm{\Omega})$ and
\begin{equation}
  \tilde{\bar{R}}^{(P,SIC)}_0(\bm{\Psi},\bm{\Omega})=\frac{1}{\log2}\left\{\nabla_0+\log|\bm{\Omega}|-\tr[\bm{\Omega}(\bm{\Psi}+\sigma^2\qI_L)]+L\right\}.
\end{equation}
Assume that $\bm{\Psi}^{\star}$ is the optimal result. Now, for problem \eqref{appr_subpro1 problem} the optimal value of $\bm{\Omega}$, according to \textbf{Lemma 5}, is
\begin{equation}
 \bm{\Omega}^{\ast}=(\bm{\Psi}^{\star}+\sigma^2\qI_L)^{-1}.
\end{equation}
 We observe that $\tilde{\bar{R}}^{(P,SIC)}(\bm{\Psi},\bm{\Omega})$ is nonconvex over both $\bm{\Psi}$ and $\bm{\Omega}$. However, it is convex over either $\bm{\Psi}$ or $\bm{\Omega}$ when the other one is fixed. Besides, the fronthaul constraint $\bar{R}^{(fh,\kappa)}(\bm{\Psi})$ is convex over  $\bm{\Psi}$, no matter what compression scheme is adopted. Therefore, an iterative coordinate ascent algorithm is proposed to find solutions to problem \eqref{appr_subpro1 problem}, as shown in \textbf{Algorithm \ref{algor1}}.

 Note that  the objective function of problem \eqref{appr_subpro1 problem} is continuously differentiable  and  the feasible set is nonempty, closed, and convex. Furthermore, the iterates $(\bm{\Psi}^{(t)},\bm{\Omega}^{(t)})$ of \textbf{Algorithm \ref{algor1}} are bounded due to the fronthaul capacity constraint. \textbf{Algorithm \ref{algor1}} yields a nondecreasing sequence of objective values for problem  \eqref{appr_subpro1 problem}. According to \cite[Corollary 2]{GRIPPO200On}, it can be verified that \textbf{Algorithm \ref{algor1}} converges to a stationary point of problem  \eqref{appr_subpro1 problem}. Given that $(\bm{\Psi}^{\star},\bm{\Omega}^{\star})$ is a stationary point of problem \eqref{appr_subpro1 problem}, then we have
 \begin{equation}\label{gradient condi1}
  \tr\left(\nabla_{\bm{\Psi}}\tilde{\bar{R}}^{(P,SIC)}(\bm{\Psi}^{\star},\bm{\Omega}^{\star})^H(\bm{\Psi}-\bm{\Psi}^{\star})\right)\leq 0, \forall \bm{\Psi}\in \mathcal{W},
\end{equation}
 where $\mathcal{W}=\{\bm{\Psi}|\eqref{cond5},\ \eqref{cond3}, \ \text{and} \ \eqref{cond4} \} $ and $\nabla_{\bm{\Psi}}\tilde{\bar{R}}^{(P,SIC)}(\bm{\Psi}^{\star},\bm{\Omega}^{\star})$ denotes the gradient of $\tilde{\bar{R}}^{(P,SIC)}(\bm{\Psi}^{\star},\bm{\Omega}^{\star})$ with respect to $\bm{\Psi}$. Substituting $\bm{\Omega}^{\ast}=(\bm{\Psi}^{\star}+\sigma^2\qI_L)^{-1}$ into \eqref{gradient condi1}, then it can be verified
$\nabla_{\bm{\Psi}}\bar{R}^{(P,SIC)}(\bm{\Psi}^{\star})=\nabla_{\bm{\Psi}}\tilde{\bar{R}}^{(P,SIC)}(\bm{\Psi}^{\star},(\bm{\Psi}^{\star}+\sigma^2\qI_L)^{-1}).$
Therefore, $(\bm{\Psi}^{\star},\bm{\Omega}^{\star})$ is  also a stationary point of problem \eqref{subpro11}.

\begin{algorithm}[htb]
\caption{Optimal algorithm for problem \eqref{appr_subpro1 problem}.}
\label{algor1}
\begin{algorithmic}[1]
\REQUIRE Large-scale channel coefficients and the tolerance factor $\epsilon$.
\ENSURE $\bm{\Psi}$.
\STATE \textbf{Initialization:} $\bm{\Psi}^{(0)}=\sigma^2\qI_L$, $\bm{\Omega}^{(0)}=(\bm{\Psi}^{(0)}+\sigma^2\qI_L)^{-1}$, and $t=1$. Calculate $\tilde{\bar{R}}^{(P,SIC,0)}$. \\
\STATE Given $\bm{\Omega}^{(t-1)}$, find the optimal $\bm{\Psi}^{(t)}$ via solving problem \eqref{appr_subpro1 problem}.\\
\STATE Update $\bm{\Omega}^{(t)}=(\bm{\Psi}^{(t)}+\sigma^2\qI_L)^{-1}$ and $\tilde{\bar{R}}^{(P,SIC,t)}$.\\

\IF{$(|\tilde{\bar{R}}^{(P,SIC,t)}(\bm{\Psi}^{(t)},\bm{\Omega}^{(t)})-\tilde{\bar{R}}^{(P,SIC,t-1)}(\bm{\Psi}^{(t-1)},\bm{\Omega}^{(t-1)})|>\epsilon)$ }
\STATE $t=t+1$ and go to Step 2;
\ELSE
\RETURN $\bm{\Psi}^{(t)}$.
\ENDIF
\end{algorithmic}
\end{algorithm}

\subsection{Optimization of Bandwidth Allocation}\label{op of bandwidth eta}
Once the optimal value of $\bm{\Psi}$ is obtained from problem \eqref{appr_subpro1 problem}, it can be used in the following optimization problem of bandwidth allocation:
\begin{subequations}
\begin{align}
\mathop{\max}_{\eta} \quad &  \bar{R}^{(B,SIC)}(\eta)+\bar{R}^{(P,SIC)}(\eta)\label{subpro2}\\
   \text{s.t.}\quad & \eqref{cond5} \quad\text{and}\quad \eqref{cond2},
\end{align}
\end{subequations}
which is a linear problem over $\eta$. Therefore, the optimal value of $\eta$ is given as
\begin{equation}\label{opeta}
  \eta^{\ast}=\frac{C_0}{\bar{R}^{(fh,\kappa)}_0+C_0}.
\end{equation}

Note that the  bandwidth allocation factor $\eta$ increases as the normalized fronthaul capacity $C_0$ rises, typically when $C_0\rightarrow\infty$, $\eta\rightarrow 1$. That is to say, all bandwidth should be used in radio access networks if there is no limitation on fronthaul. However, as $C_0\rightarrow 0$, suggesting that the unit fronthaul capacity is small, then more bandwidth should be allocated to fronthaul links when fronthaul compression noise is fixed.

\subsection{Joint Optimization of $\bm{\Psi}$ and $\eta$}
By substituting the optimal value of $\eta$ given by \eqref{opeta} into \eqref{asym objec function}, we reformulate an equivalent problem of  problem \eqref{sym objec problem} as
\begin{subequations}\label{joint op of asym problem}
\begin{align}
  \mathop{\max}_{\bm{\Psi}} \quad &  \frac{\bar{R}^{(B,SIC)}_0(\bm{\Psi})+\bar{R}_0^{(P,SIC)}(\bm{\Psi})}{\bar{R}^{(fh,\kappa)}_0(\bm{\Psi})+C_0}\label{joint op of asym}\\
  \text{s.t.} \quad &\eqref{cond3} \quad \text{and} \quad \eqref{cond4},
\end{align}
\end{subequations}
which is a fractional programming whose numerator and denominator are both convex with respect to $\bm{\Psi}$. For the convenience of designing the related algorithm, we apply the inequality \eqref{inequality} again to the numerator of \eqref{joint op of asym}, thus problem \eqref{joint op of asym problem} becomes
\begin{subequations}\label{joint op of asym2 problem}
\begin{align}
\mathop{\max}_{\bm{\Psi},\bm{\Omega}} \quad   &\frac{f(\bm{\Psi},\bm{\Omega})}{g(\bm{\Psi})}\label{joint op of asym2}\\
\text{s.t.} \quad &\eqref{cond3} \quad \text{and} \quad \eqref{cond4},
\end{align}
\end{subequations}
where $f(\bm{\Psi},\bm{\Omega})=\bar{R}^{(B,SIC)}_0+\tilde{\bar{R}}_0^{(P,SIC)}(\bm{\Psi},\bm{\Omega})$ and $g(\bm{\Psi})=\bar{R}^{(fh,\kappa)}_0(\bm{\Psi})+C_0$. Therefore, the object function in  \eqref{joint op of asym2} is a concave-convex fractional function over $\bm{\Psi}$ but a concave function over $\bm{\Omega}$, respectively. Before the algorithm to find the optimal solutions is presented, we first define the set $\calS=\{\bm{\Psi}|\bm{\Psi}_{ll}\geq 0,\bm{\Psi}_{ll^{\prime}}=0,l\neq l^{\prime}, l=1,\ldots,L\}$ to describe constraints \eqref{cond3} and \eqref{cond4} and introduce the following proposition:
\begin{proposition}\label{proposition}
Assume that $f(\bm{\Psi},\bm{\Omega})$ and $g(\bm{\Psi})$ are continuous, and $f(\bm{\Psi},\bm{\Omega})$ is positive. Given $\bm{\Psi}^{\ast}\in\mathcal{S}$, $\bm{\Omega}^{\ast}=(\bm{\Psi}^{\ast}+\sigma^2\qI_L)^{-1}$, and $\omega^{\ast}=\frac{f(\bm{\Psi}^{\ast},\bm{\Omega}^{\ast})}{g(\bm{\Psi}^{\ast})}$. Then  $\bm{\Psi}^{\ast}\in\mathcal{S}$ is the optimal solution to problem \eqref{joint op of asym2 problem} if and only if
\begin{equation}
  (\bm{\Psi}^{\ast},\bm{\Omega}^{\ast})=\mathop{\argmax}_{\bm{\Psi}\in\mathcal{S},\bm{\Omega}}[f(\bm{\Psi},\bm{\Omega})-\omega^{\ast}g(\bm{\Psi})].
\end{equation}
\end{proposition}
\IEEEproof See Appendix \ref{appendixB} for reference.

\textbf{Proposition} \ref{proposition} suggests that we can find the optimal solutions to fractional problem \eqref{joint op of asym2 problem} by finding the unique zero of the auxiliary function $F(\omega)=f(\bm{\Psi},\bm{\Omega})-\omega g(\bm{\Psi})$. Therefore, we propose an algorithm based on Dinkelbach's algorithm \cite{dinkelbach1967nonlinear,zappone2015energy}, as shown in \textbf{Algorithm} \ref{algor2}.

\begin{algorithm}[htb]
\caption{Joint optimization algorithm for problem \eqref{joint op of asym2 problem}.}
\label{algor2}
\begin{algorithmic}[1]
\REQUIRE Large-scale channel coefficients and the tolerance factors $\epsilon_1$ and $\epsilon_2$.
\ENSURE $\bm{\Psi}$.
\STATE \textbf{Initialization:} $\underline{\bm{\Psi}}^{(0)}\!=\!\sigma^2\qI_L$, $\bm{\Omega}^{(0)}=(\underline{\bm{\Psi}}^{(0)}+\sigma^2\qI_L)^{-1}$, $\omega^{(0)}$=0, and $t=0$.
\WHILE{$F(\omega^{(t)})>\epsilon_1$}
\STATE  $G^{(0)}=f(\bm{\Psi}^{(0)},\bm{\Omega}^{(0)})-\omega^{(t)} g(\bm{\Psi}^{(0)})$.
\STATE $i=1$;
\STATE $\underline{\bm{\Psi}}^{(i)}
=\mathop{\argmax}_{\bm{\Psi}\in\mathcal{S}}\{f(\bm{\Psi},\bm{\Omega}^{(i-1)})-\omega^{(t)} g(\bm{\Psi})\}$.
\STATE Update $\bm{\Omega}^{(i)}=(\underline{\bm{\Psi}}^{(i)}+\sigma^2\qI_L)^{-1}$.
\STATE $G^{(i)}=f(\underline{\bm{\Psi}}^{(i)},\bm{\Omega}^{(i)})-\omega^{(t)} g(\underline{\bm{\Psi}}^{(i)})$.
\IF{$|G^{(i)}-G^{(i-1)}|>\epsilon_2$ }
\STATE $i=i+1$ and go to Step 5;
\ELSE
\STATE $\bm{\Psi}^{(t)}=\underline{\bm{\Psi}}^{(i)}$.
\ENDIF
\STATE $F(\omega^{(t)})=f(\bm{\Psi}^{(t)},\bm{\Omega}^{(t)})-\omega^{(t)} g(\bm{\Psi}^{(t)})$.
\STATE $\omega^{(t+1)}=\frac{f(\bm{\Psi}^{(t)},\bm{\Omega}^{(t)})}{g(\bm{\Psi}^{(t)})}$.
\STATE $t=t+1$.
\ENDWHILE
\RETURN $\bm{\Psi}^{(t-1)}$.
\end{algorithmic}
\end{algorithm}

\textbf{Algorithm} \ref{algor2}  is a two-level iterative algorithm. The inner iteration aims to make $F(\omega)$ converge to a certain value  with a given $\omega$ and the outer iteration tries to find a typical value $\omega^{\ast}$ which establishes the equality $F(\omega^{\ast})=0$.

The convergence of \textbf{Algorithm} \ref{algor2}  is given by the following proposition.
\begin{proposition}
\textbf{Algorithm} \ref{algor2} converges to the solution of \textbf{Proposition} \ref{proposition}.
\end{proposition}
\IEEEproof Refer to Appendix \ref{AppendixC}.

Since \textbf{Algorithm} \ref{algor2} translates the original concave-convex problem into a sequence of auxiliary subproblems indexed by the parameter $\omega$, the overall computational complexity relies on the complexities of all subproblems and the convergence rate of the subproblem sequence.
To explain the convergence of subproblem sequence, we rewritten \eqref{update rule} as
\begin{equation}\label{convergence}
  \omega^{(t+1)}=\omega^{(t)}-\frac{F(\omega^{(t)})}{F^{\prime}(\omega^{(t)})},
\end{equation}
where $F^{\prime}(\omega^{(t)})=\frac{\partial F(\omega^{(t)})}{\partial \omega^{(t)}}$. Equation \eqref{convergence} can be interpreted as an application of Newton's method to $F(\omega)$ so that the subproblem sequence in \textbf{Algorithm} \ref{algor2} converges at a super-linear rate \cite{zappone2015energy}.

\section{Joint Design under Linear Reception without SIC}\label{analysis2}
In the case of the linear reception without SIC, the joint optimization problem is formulated as:
\begin{subequations}\label{asym objec problem with mmse}
\begin{align}
  \mathop{\max}_{\eta,\bm{\Psi}} \quad &  \bar{R}^{(B,L)}(\eta)+ \bar{R}^{(P,L)}(\eta,\bm{\Psi})\label{asym objec function with mmse}\\
   \text{s.t.}\quad &\eqref{cond2}-\eqref{cond4} \ \text{and} \ \eqref{cond5}.
\end{align}
\end{subequations}
Similar to the analysis on problem \eqref{joint op of asym problem} whose achievable SR is based on the linear reception with SIC, we translate problem \eqref{asym objec problem with mmse} into a concave-convex fractional programming formulated as:
\begin{subequations}\label{asym object problem with lmmse2}
\begin{align}
  \mathop{\max}_{\bm{\Psi},\{\bm{\Omega}_m\}_{m=1}^{M}} \quad &  \frac{\bar{R}^{(B,L)}_0+\tilde{\bar{R}}_0^{(P,L)}(\bm{\Psi},\{\bm{\Omega}_m\}_{m=1}^{M})}{\bar{R}^{(fh,\kappa)}_0(\bm{\Psi})+C_0}\label{asym object func with lmmse2}\\
   \text{s.t.}\quad &\eqref{cond3}\quad \text{and} \quad \eqref{cond4},
\end{align}
\end{subequations}
where $\tilde{\bar{R}}_0^{(P,L)}(\bm{\Psi},\{\bm{\Omega}_m\}_{m=1}^{M})=\sum\limits_{m=1}^M{\tilde{\bar{R}}_{0,m}^{(P,L)}}(\bm{\Psi},\bm{\Omega}_m)$ and
\begin{align}
\tilde{\bar{R}}^{(P,L)}_{0,m}(\bm{\Psi},\bm{\Omega}_m)&=\frac{1}{\log2}\Bigg[\nabla_0+\log|\bm{\Omega}_m|-\tr(\bm{\Omega}_m\bm{\tilde{\Lambda}}_m)+L \nonumber\\
&-\sum\limits_{m^{\prime}=1,m^{\prime}\neq m}^M{(\frac{1}{1+b_{mm^{\prime}}}-\log\frac{1}{1+b_{mm^{\prime}}})}-(M-1)\Bigg].
\end{align}
Actually, problem \eqref{asym object problem with lmmse2} can also be solved via \textbf{Algorithm \ref{algor2}} with some modification as follows:

\begin{itemize}
  \item   Replace auxiliary function $f(\bm{\Psi}, \bm{\Omega})$ with a new defined auxiliary function
\begin{equation}
  \bar{f}(\bm{\Psi}, \{\bm{\Omega}_m\}_{m=1}^{M})=R^{(B,L)}_{0}+\sum_{m=1}^M{\tilde{\bar{R}}_{0,m}^{(P,L)}(\bm{\Psi}, \bm{\Omega}_m)}. \nonumber
\end{equation}
  \item Update (initialize) $\bm{\Omega}_m^{(i)}=(\sum\limits_{m^{\prime}=1,m^{\prime}\neq m}^M{\frac{p_{m^{\prime}}\qU_{m^{\prime}}}{1+\tilde{b}_{mm^{\prime}}}+\underline{\bm{\Psi}}^{(i)}+\sigma^2\qI_L})^{-1}$, for $m=1,\ldots,M$.
\end{itemize}

\section{Suboptimal Analysis}\label{subanalysis}
The Karush-Kuhn-Tucker (KKT) condition is a necessary condition of the optimal solutions to problems \eqref{sym objec problem} and \eqref{asym objec problem with mmse}. To get the KKT condition, we build the  Lagrangian function associated with such two problems as
\begin{align}
  L(\bm{\Psi},\bm{\lambda},\qv)=\eta[\bar{R}^{(B,\varsigma)}_0+\bar{R}^{(P,\varsigma)}_0(\bm{\Psi})]-\lambda_1[\eta\bar{R}^{(fh,\kappa)}_0(\bm{\Psi})-(1-\eta)C_0]\nonumber+\sum\limits_{l=1}^L{v_l\psi^2_l}+\lambda_2\eta-\lambda_3(\eta-1),
\end{align}
where $\bm{\lambda}=[\lambda_1, \lambda_2, \lambda_3]^T$ and $\qv=[v_1,\ldots,v_L]^T$ are vectors composed of the Lagrangian multipliers.
Let $\partial L\setminus \partial \psi_l^2=0$, then we have
 \begin{equation}\label{optimality condition}
   (A_l+\psi^2_l+\sigma^2)^{-1}-(B^{(\varsigma)}_l+\psi^2_l+\sigma^2)^{-1}-\lambda_1[(A^{(\kappa)}_l+\psi^2_l+\sigma^2)^{-1}-(\psi^2_l)^{-1}]=0,
 \end{equation}
where $v_l=0$ is omitted, $A_l=\sum_{m=1}^M{\frac{p_m\mu_{lm}}{1+b_m}}$,
\begin{equation}
A^{(\kappa)}_l=
\begin{cases}
A_l,  \kappa=\text{WZ},\\
\sum\limits_{m=1}^M{p_m\mu_{lm}}, \kappa=\text{P2P},
\end{cases}
\end{equation}
and
\begin{equation}
B^{(\varsigma)}_l=
\begin{cases}
0, \text{$\varsigma$=SIC},\\
\sum\limits_{m=1}^M\left(\sum\limits_{m^{\prime}=1,m^{\prime}\neq m}^M{\frac{p_{m^{\prime}} u_{lm^{\prime}}}{1+\tilde{b}_{mm^{\prime}}}}\right), \text{$\varsigma$=L}.
\end{cases}
\end{equation}

\subsection{Small $\eta$-value Case}
A small $\eta$ value implies that  more bandwidth is allocated to fronthaul links, then the quantization degree can be reduced. Thus the values of the diagonal elements of matrix $\bm{\Psi}$, $\psi_l^2$'s, become small, leading to a high SQNR.

Consider the case where the linear reception with SIC is adopted, i.e., $\varsigma$=SIC. As analyzed above, when $\eta$ value is very small, the compression noise is also very small and  $A^{(\kappa)}_l\gg(\psi^2_l+\sigma^2)$. Under this high SQNR condition, we have
\begin{equation}
 (A^{(\kappa)}+\psi^2_l+\sigma^2)^{-1}\ll(\psi^2_l+\sigma^2)^{-1},
\end{equation}
therefore, the optimality condition \eqref{optimality condition} becomes
\begin{equation}
  \psi_l^2\approx\frac{\lambda_1}{1-\lambda_1}\sigma^2,\label{appro deter of psi}
\end{equation}
where $\lambda_1\in[0,1)$. Note that \eqref{appro deter of psi} does not hold when $\varsigma$=L. \eqref{appro deter of psi} indicates that under linear reception with SIC, whatever compression scheme (either WZ coding or P2P compression) is applied at the RRHs, the optimal compression noise variance should be proportional to the background noise level when $\eta$  value is small. This conclusion also suggests a simple way, referred to as uniform quantization scheme where each RRH has the same quantization noise variance, to find suboptimal solutions to problems \eqref{appr_subpro1 problem}, \eqref{joint op of asym problem} and \eqref{asym objec problem with mmse}. The details are given as follows.

We assume $\bm{\Psi}=\varrho\sigma^2\qI_L$. For \eqref{appr_subpro1 problem}, the fronthaul capacity constraint becomes a monotonically decreasing function over $\varrho$, thus we just need to find a certain $\varrho^{\ast}$, which satisfies
\begin{equation}
  \bar{R}^{(fh,\kappa)}(\varrho^{\ast})=(1-\eta)C_0.
\end{equation}
Then $\bm{\Psi}=\varrho^{\ast}\sigma^2\qI_L$ is the output. The suboptimal algorithm with uniform quantization for problem \eqref{appr_subpro1 problem} as shown in \textbf{Algorithm \ref{algor3}} has lower  complexity than \textbf{Algorithm \ref{algor1}}. Actually, under certain conditions,  \textbf{Algorithm \ref{algor3}} can achieve near optimal performance. More discussion is presented in Section \ref{numerical results}.

With the assumption $\bm{\Psi}=\varrho\sigma^2\qI_L$, problems \eqref{joint op of asym problem} and \eqref{asym objec problem with mmse} can be  simplified to a single-variable problem of $\varrho$,
\begin{subequations}
\begin{align}
  \mathop{\max}_{\varrho} \quad &  \frac{\bar{R}^{(B,\varsigma)}_0+\bar{R}_0^{(P,\varsigma)}(\varrho)}{\bar{R}^{(fh,\kappa)}_0(\varrho)+C_0}\\
   \text{s.t.}\quad &0\leq\varrho<1,
\end{align}
\end{subequations}
which can be  solved via one-dimension search.

\begin{algorithm}[htb]
\caption{Uniform quantization algorithm for problem \eqref{appr_subpro1 problem}.}
\label{algor3}
\begin{algorithmic}[1]
\REQUIRE Large-scale channel coefficients, step factor $\vartheta>1$, and the tolerance factor $\epsilon$.
\ENSURE $\bm{\Psi}$.
\STATE \textbf{Initialization:} $\varrho^{(0)}=10^{-10}$, $\bm{\Psi}^{(0)}=\varrho^{(0)}\sigma^2\qI_L$, and $t=0$. Calculate $\bar{R}^{(fh,\kappa,0)}$. \\
\WHILE{$(\eta\bar{R}^{(fh,\kappa,t)}-(1-\eta)C_0>0)$}
\STATE $\varrho_{min}^{(t)}=\varrho^{(t)}$ and $\varrho_{max}^{(t)}=\vartheta\varrho^{(t)}$; $t=t+1$.
\STATE Update $\varrho^{(t)}=\varrho_{max}^{(t-1)}$ and $\bm{\Psi}^{(t)}=\varrho^{(t)}\sigma^2\qI_L$.\\
\STATE Recalculate $\bar{R}^{(fh,\kappa,t)}$ with $\bm{\Psi}^{(t)}$.\\
\ENDWHILE
\STATE Find the optimal $\varrho^{\star}$ in [$\varrho_{min}^{(t)}$, $\varrho_{max}^{(t)}$] via one-dimensional search until $|\eta\bar{R}^{(fh,\kappa,t)}-(1-\eta)C_0|<\epsilon$.

\RETURN $\bm{\Psi}=\varrho^{\star}\sigma^2\qI_L$.
\end{algorithmic}
\end{algorithm}

\subsection{Large $\eta$-value case}
A large $\eta$ value suggests that less bandwidth is allocated to fronthaul links, so the RRHs have to increase the quantization level to meet the fronthaul capacity constraint,  resulting in the larger values of $\{\psi^2_l\}_{l=1}^{L}$. In the case where WZ coding is adopted, i.e., $\varsigma$=WZ,  when  $\eta$ is large, the diagonal elements of $\bm{\Psi}$ are so large that we can ignore the effect of background noise, thus
\begin{equation}
  \bar{R}^{(P,SIC)}(\eta,\bm{\Psi})\approx\bar{R}^{(fh,WZ)}(\eta,\bm{\Psi})=(1-\eta)C_0,
\end{equation}
which holds regardless of what kind of WZ coding, optimal compression or uniform compression, is adopted. Therefore, when $\eta$ is large enough and the linear reception with SIC is adopted, WZ coding with uniform compression can achieve near optimal performance.

\textbf{Remark:} Based on the aforementioned analysis, since WZ coding with uniform compression  can achieve near optimal performance in both small $\eta$-value and large $\eta$-value regimes when the linear reception with SIC is adopted, we further draw a conclusion that if the linear reception with SIC  is adopted, WZ coding with uniform compression can achieve the performance near to that of WZ coding with optimal compression.

\section{Numerical Results}\label{numerical results}
In this section, we first validate the accuracy of the deterministic approximations for the ergodic SRs, and then use these deterministic results to investigate  the system performance under different compression strategies. We consider a simple system where a macro BS with $N=10$ antennas is deployed at the center, radiating over an area with a radius of 250 m. We assume the locations of RRHs and MUEs are determined respectively according to uniform distribution  and  SUEs are distributed uniformly within the coverage of RRHs with a radius of 50 m. The transmit power of all UEs is given as $p^{(M)}=p^{(S)}=23$ dBm, and the noise power spectral density is $-174$ dBm/Hz.  It is assumed that the path-loss model is described as $31.5+40\log_{10}(d)$ for the SUE-to-RRH and MUE-to-RRH links and  $31.5+35\log_{10}(d)$ for the MUE-to-BS link, respectively, where $d$ in meters denotes the distance \cite{Peng2015Energy}. The available spectrum resource is $F=20$ MHz, which is allocated by the BS between the radio access networks and fronthaul links.
\subsection{Validation of Accuracy of Deterministic Equivalents}
\begin{figure}
\includegraphics[width=0.6\textwidth]{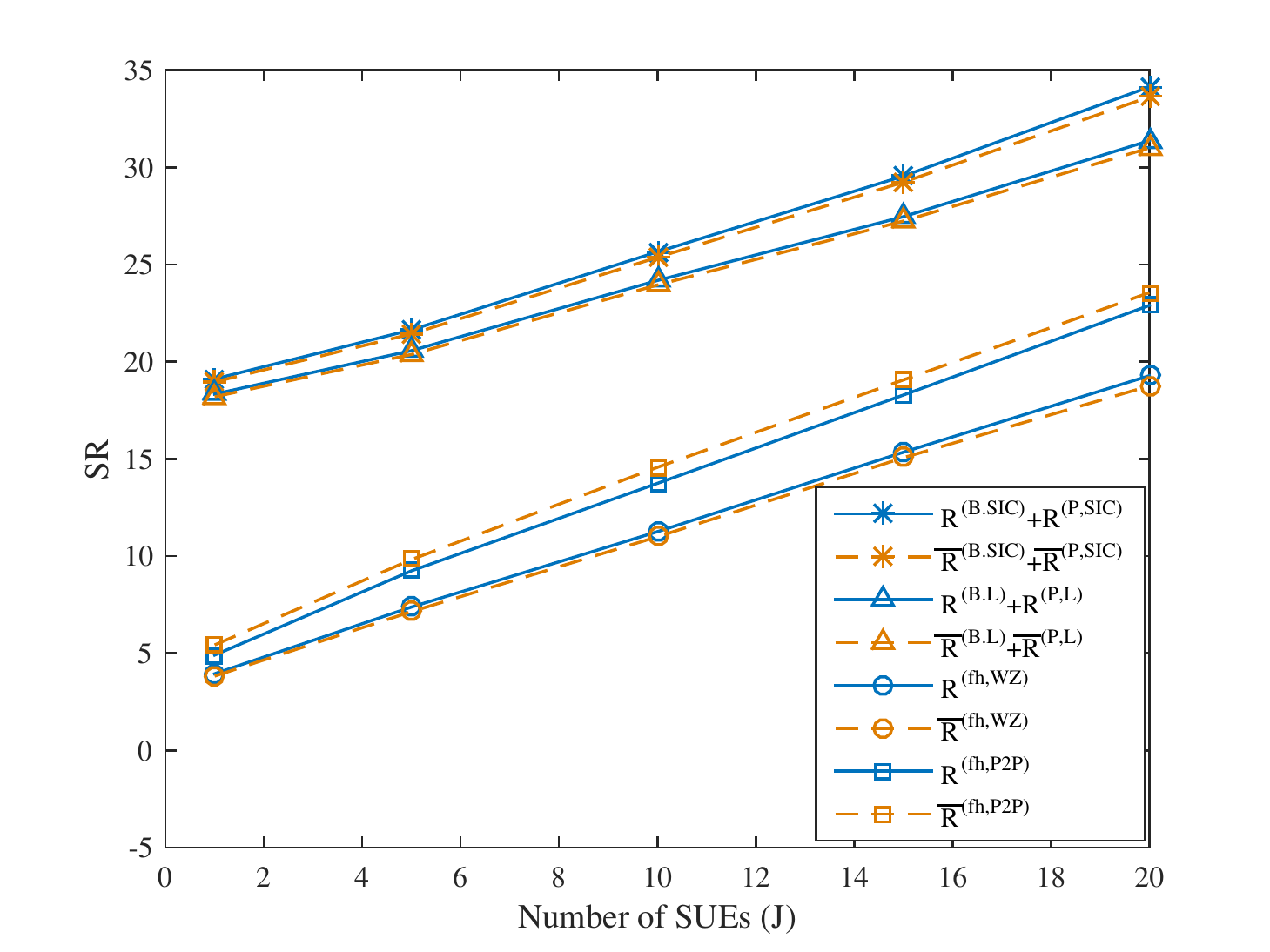}
\centering
\caption{Comparison of ergodic SRs and their deterministic approximations versus the number of SUEs $J$ with $\{N=10, K=5, L=30, \bm{\Psi}=10^{-10}\qI_L, \text{and} \ \eta=0.5 \}$.} \label{ergo_deter_versus_J}
\end{figure}

We first compare  Monte Carlo simulation results $R^{(B,\varsigma)}+R^{(P,\varsigma)}$  and  $R^{(fh,\kappa)}$ under $10^3$ trials with their analytical results $\bar{R}^{(B,\varsigma)}+\bar{R}^{(P,\varsigma)}$ and $\bar{R}^{(fh,\kappa)}$ to show the accuracy of  these deterministic equivalents. We assume $N=10$, $K=5$, $L=30$, $\bm{\Psi}=10^{-10}\qI_L$, and $\eta=0.5$. It is observed from Fig. \ref{ergo_deter_versus_J} that these deterministic approximations are very close to their simulation results. Fig. \ref{ergo_deter_versus_J} also shows that the linear reception with SIC achieves a larger SR than that without SIC because of less interference under the linear reception with SIC. Besides, given a fixed $\bm{\Psi}$ value, we find that $\bar{R}^{(fh,WZ)}$ is smaller than $\bar{R}^{(fh,P2P)}$, which suggests the WZ coding scheme allows a larger quantization  noise variance  than the  P2P compression scheme when the fronthaul capacity is limited.
\subsection{Comparison with Other Schemes}
\begin{figure}
  \centering
  \subfigure[]{
    \label{maximal_sr_versus_c0}
    \includegraphics[width=0.6\textwidth]{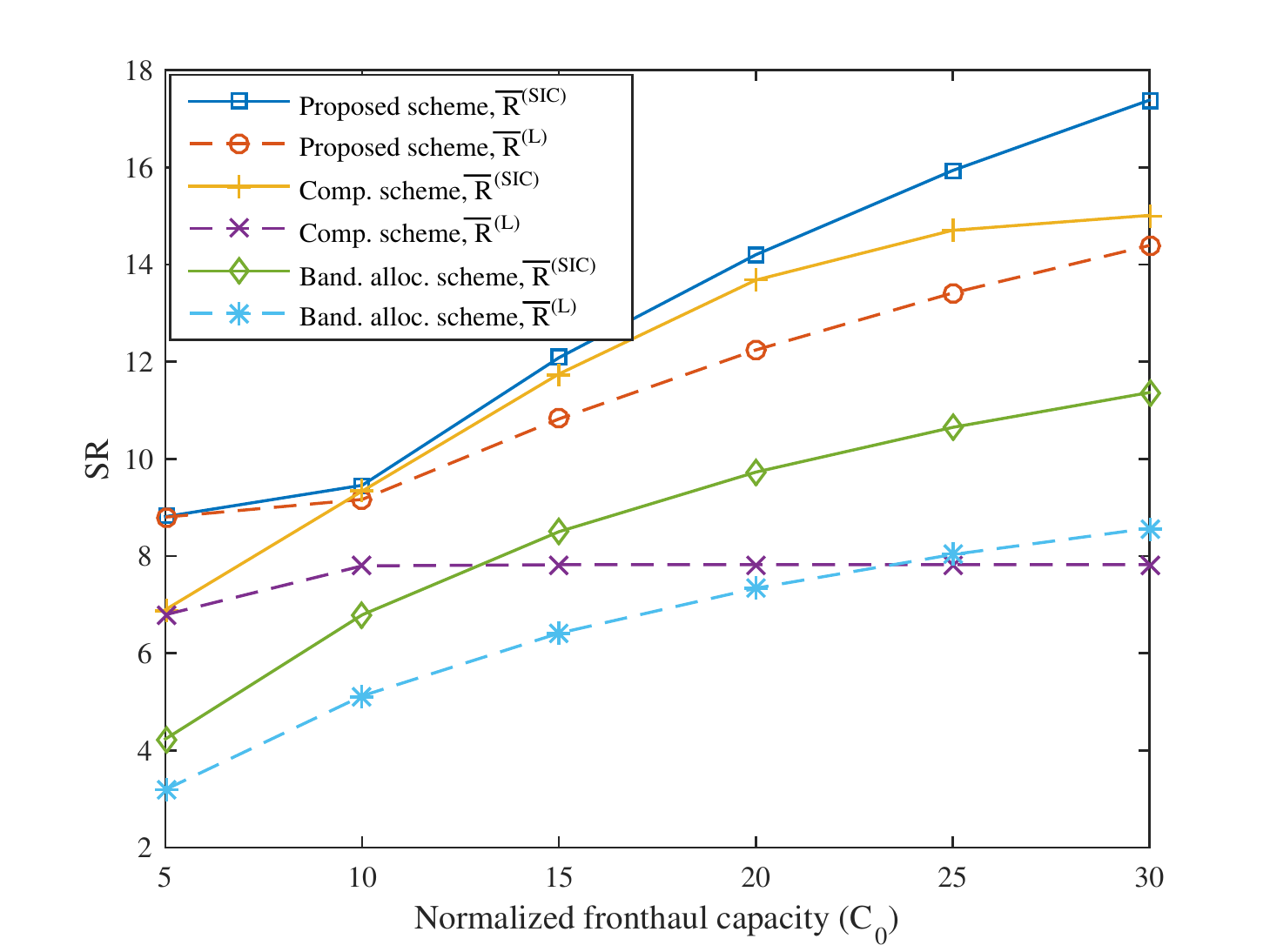}}
  \hspace{1in}
  \subfigure[]{
    \label{optimal_eta_versus_c0}
    \includegraphics[width=0.6\textwidth]{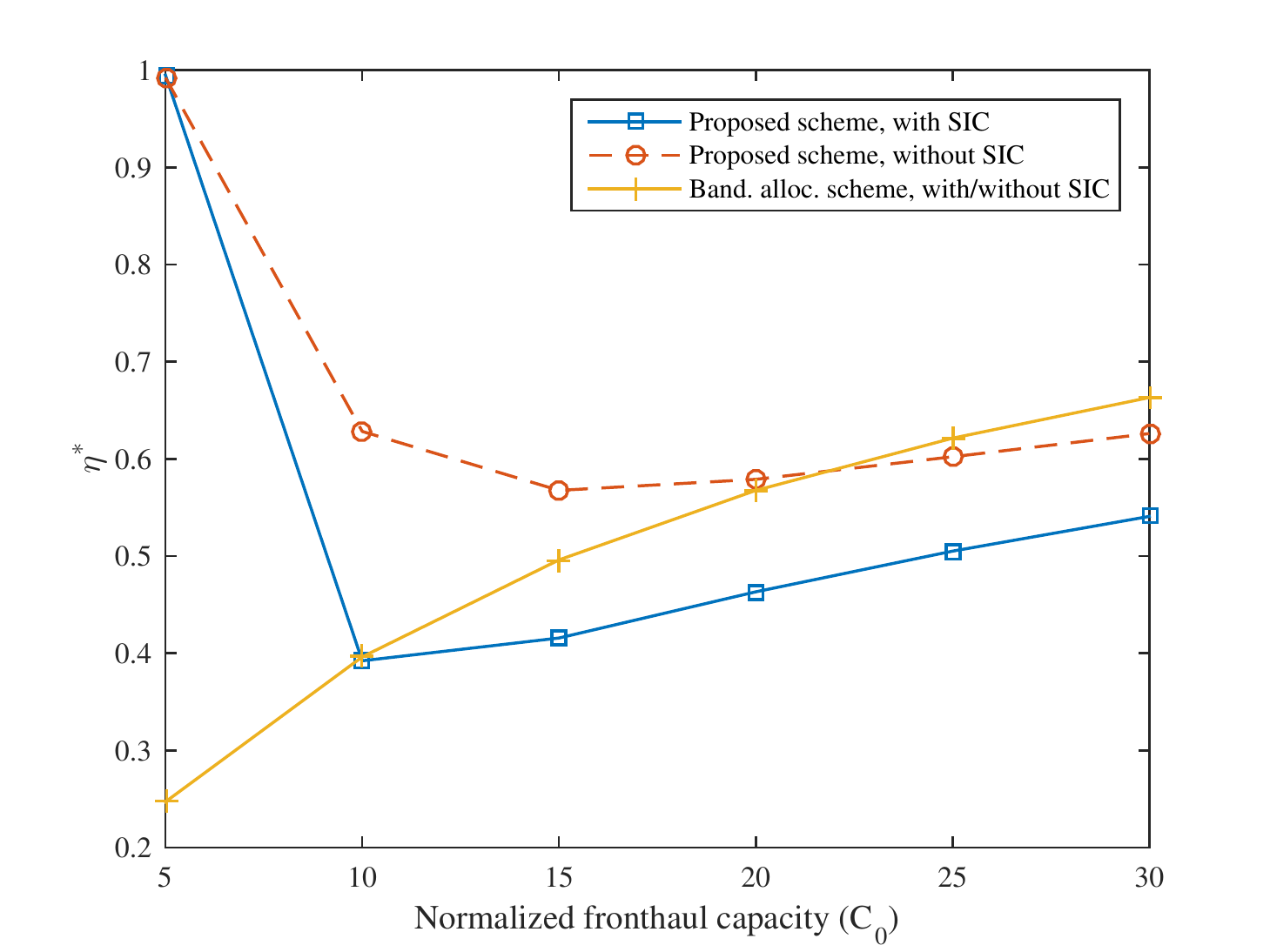}}
  \caption{Maximal achievable SR (a) and the optimal $\eta$ value (b) versus the normalized fronthaul capacity $C_0$ with \{$L=4$, $J=4$, and $K=2$\}.}
  \label{maximal_sr_optimal_eta_versus_c0} %% label for entire figure
\end{figure}
In order to show the performance of the proposed scheme, we introduce two baseline schemes: 1) Fix the bandwidth allocation factor $\eta$, then find the optimal compression noise matrix $\bm{\Psi}$ to maximize the achievable SR; and 2) fix the compression noise matrix $\bm{\Psi}$, then find the optimal bandwidth allocation factor $\eta$ to maximize the achievable SR. We refer to such two schemes as compression scheme and bandwidth allocation scheme, respectively. For convenience of expression, we define $\bar{R}^{(L)}=\bar{R}^{(B,L)}+\bar{R}^{(P,L)}$ and $\bar{R}^{(SIC)}=\bar{R}^{(B,SIC)}+\bar{R}^{(P,SIC)}$. Fig. \ref{maximal_sr_versus_c0}  depicts the achievable maximal SR versus the normalized fronthaul capacity  and Fig. \ref{optimal_eta_versus_c0} presents the corresponding optimal bandwidth allocation factor $\eta^{\ast}$. Here, we set $L=4$, $J=4$, and $K=2$ and adopt WZ coding. Besides, $\eta$ in the compression scheme (labelled as ``Comp. scheme") is set as 0.5 and $\bm{\Psi}$ in the bandwidth allocation scheme (labelled as ``Band. alloc. scheme") is set as $10^{-13}$, which is in the same order of  magnitude as the background noise.
From Fig. \ref{maximal_sr_versus_c0}, it is observed that the proposed scheme outperforms than the two baseline schemes, no matter what decoding method, linear reception with or without SIC, is adopted. This is because the proposed scheme provides more flexibility for SR maximization. We can also find that the linear reception with SIC always achieves a higher SR than that without SIC  because in the SIC scheme, the latter-decoded user messages have less interference.

Fig. \ref{optimal_eta_versus_c0} illustrates that the optimal bandwidth allocation factor $\eta^{\star}$ increases as the normalized fronthaul capacity $C_0$ rises in the bandwidth allocation scheme. Typically when $C_0\rightarrow \infty$, $\eta^{\star}\rightarrow 0$. However, the proposed scheme does not meet such a monotonously growing trend.  In the proposed scheme as $C_0$  increases, $\eta^{\star}$ also grows in the large $C_0$-value regime because the larger normalized fronthaul capacity is, the less bandwidth fronthaul needs. But in much smaller $C_0$-value regime, the fronthaul capacity is also small no matter how much bandwidth is allocated to fronthaul links, so that the achievable SR at the BBU is very small.  We aim to maximize the achievable SR including the SR at the BS and that at the BBU, thus almost all bandwidth is reserved for radio access links to maximize the SR at the BS, i.e, $\eta\approx1$.

\subsection{Comparison of WZ Coding and P2P Compression}
 \begin{figure}
  \centering
  \subfigure[]{
    \label{p2p_wz_versus_J_with_sic}
    \includegraphics[width=0.6\textwidth]{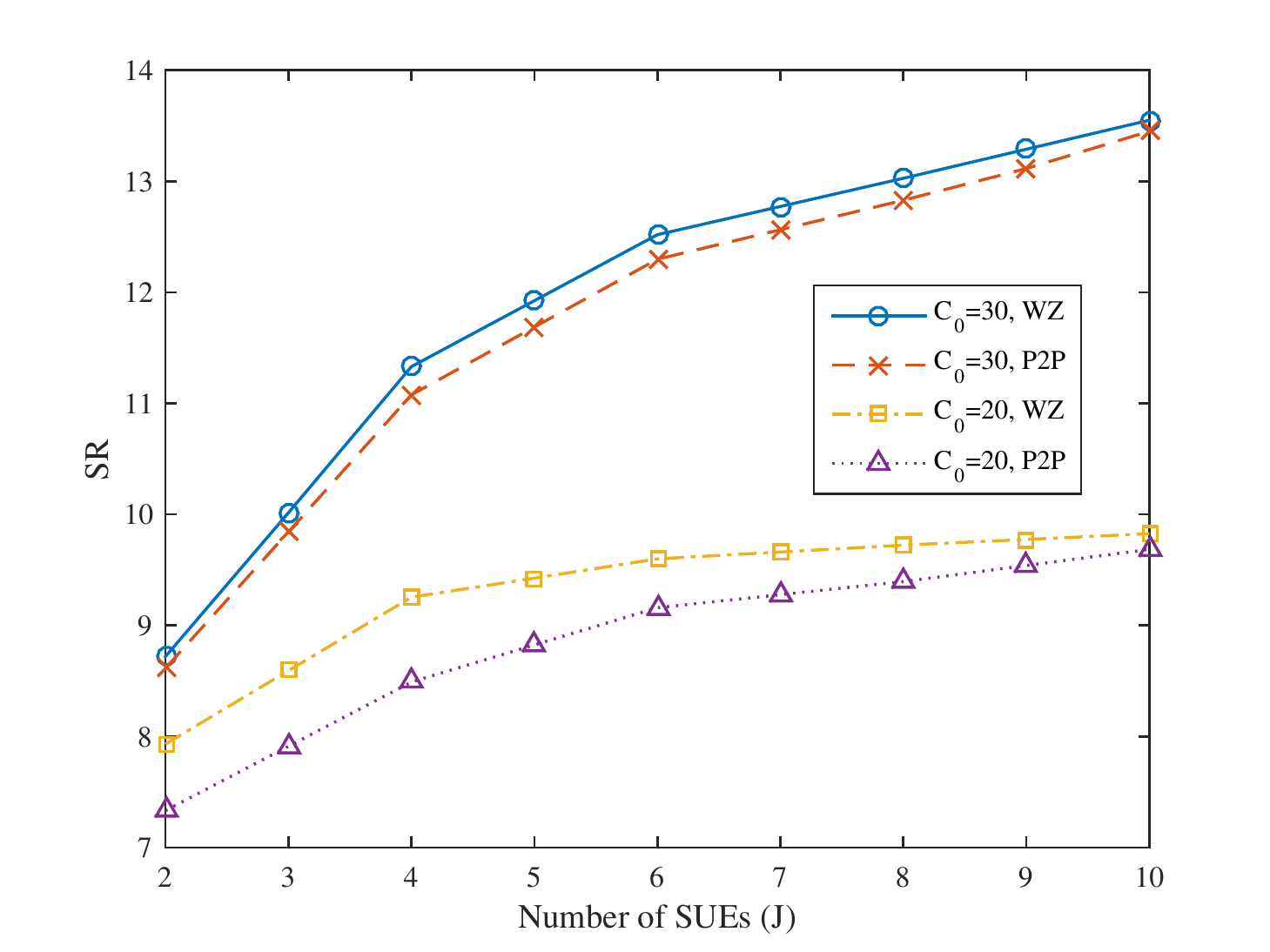}}
  \hspace{1in}
  \subfigure[]{
    \label{p2p_wz_versus_J_with_mmse}
    \includegraphics[width=0.6\textwidth]{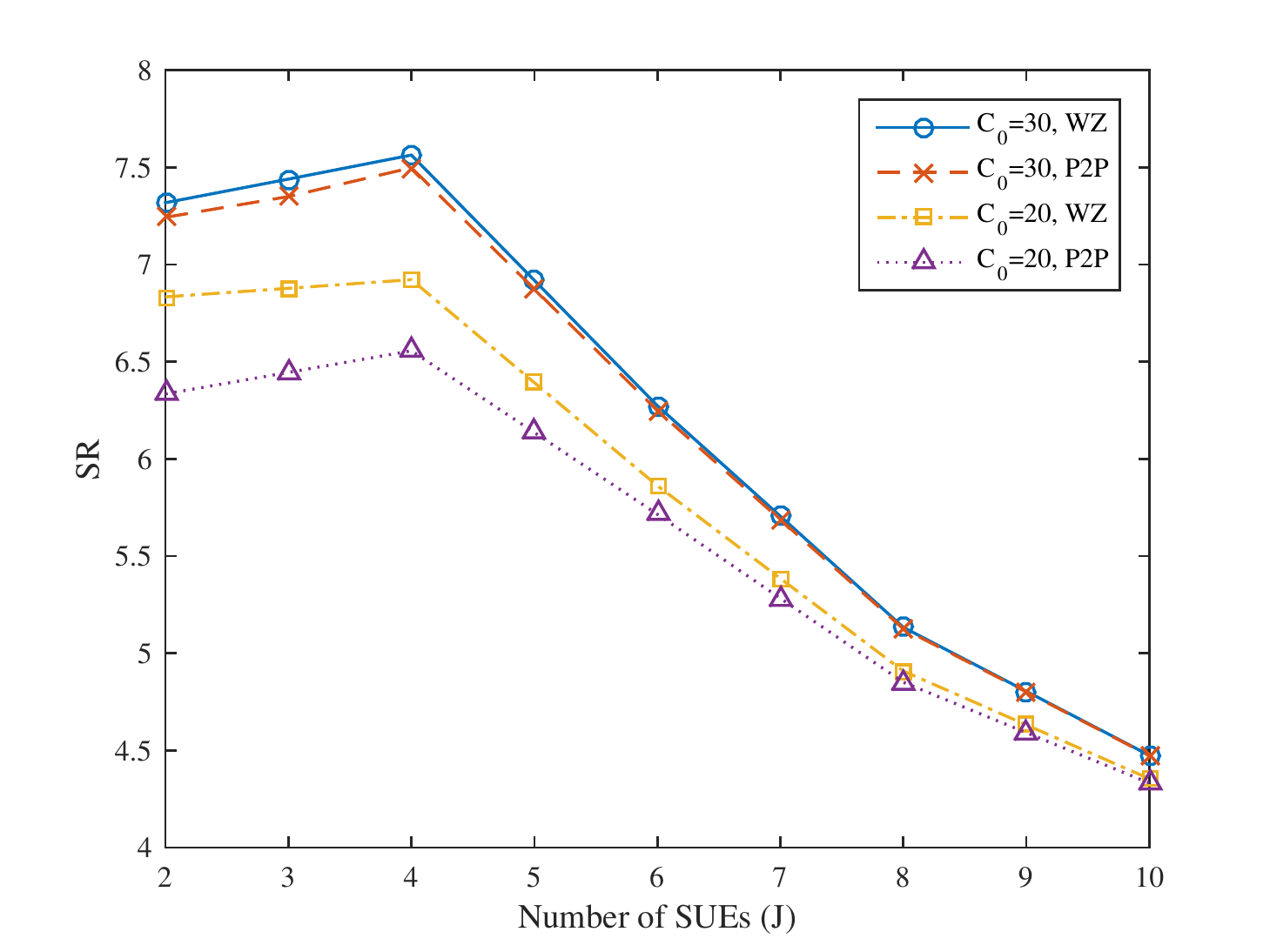}}
  \caption{(a) and (b) are the achievable SRs $\bar{R}^{(P,SIC)}$ and $\bar{R}^{(P,L)}$ versus the number of SUEs $J$ under different compression schemes, respectively, with $\{L=4, K=2, \text{and} \ \eta=0.5 \}$.}
  \label{p2p_wz_versus_J} %% label for entire figure
\end{figure}

Figs. \ref{p2p_wz_versus_J_with_sic} and \ref{p2p_wz_versus_J_with_mmse} present the achievable SRs $\bar{R}^{(P,SIC)}$ and $\bar{R}^{(P,L)}$ versus the number of SUEs $J$ under different compression schemes, respectively, with $\{L=4, K=2, \text{and} \ \eta=0.5 \}$. We observe that WZ coding always achieves a larger SR than P2P compression, because WZ coding takes advantage of the correlation of the received signals at the RRHs as side information when decompressing signals.  Figs. \ref{p2p_wz_versus_J_with_sic} and  \ref{p2p_wz_versus_J_with_mmse} also illustrate that the achievable SRs $\bar{R}^{(P,SIC)}$ and $\bar{R}^{(P,L)}$  increase with the normalized fronthaul capacity  since more bits are allowed to be transmitted to the BBU pool simultaneously.  As the number of SUEs increases, the achievable SR $\bar{R}^{(P,SIC)}$ as shown in Fig. \ref{p2p_wz_versus_J_with_sic} becomes larger.  However, the achievable $\bar{R}^{(P,L)}$  as shown in Fig. \ref{p2p_wz_versus_J_with_mmse},  first increases and then decreases because more SUEs bring more inter-user interference which hinders the growth of $\bar{R}^{(P,L)}$.

\subsection{Comparison of Uniform Compression and Optimal Compression }

\begin{figure}
  \centering
  \subfigure[]{
    \label{SR_versus_eta_wz}
    \includegraphics[width=0.6\textwidth]{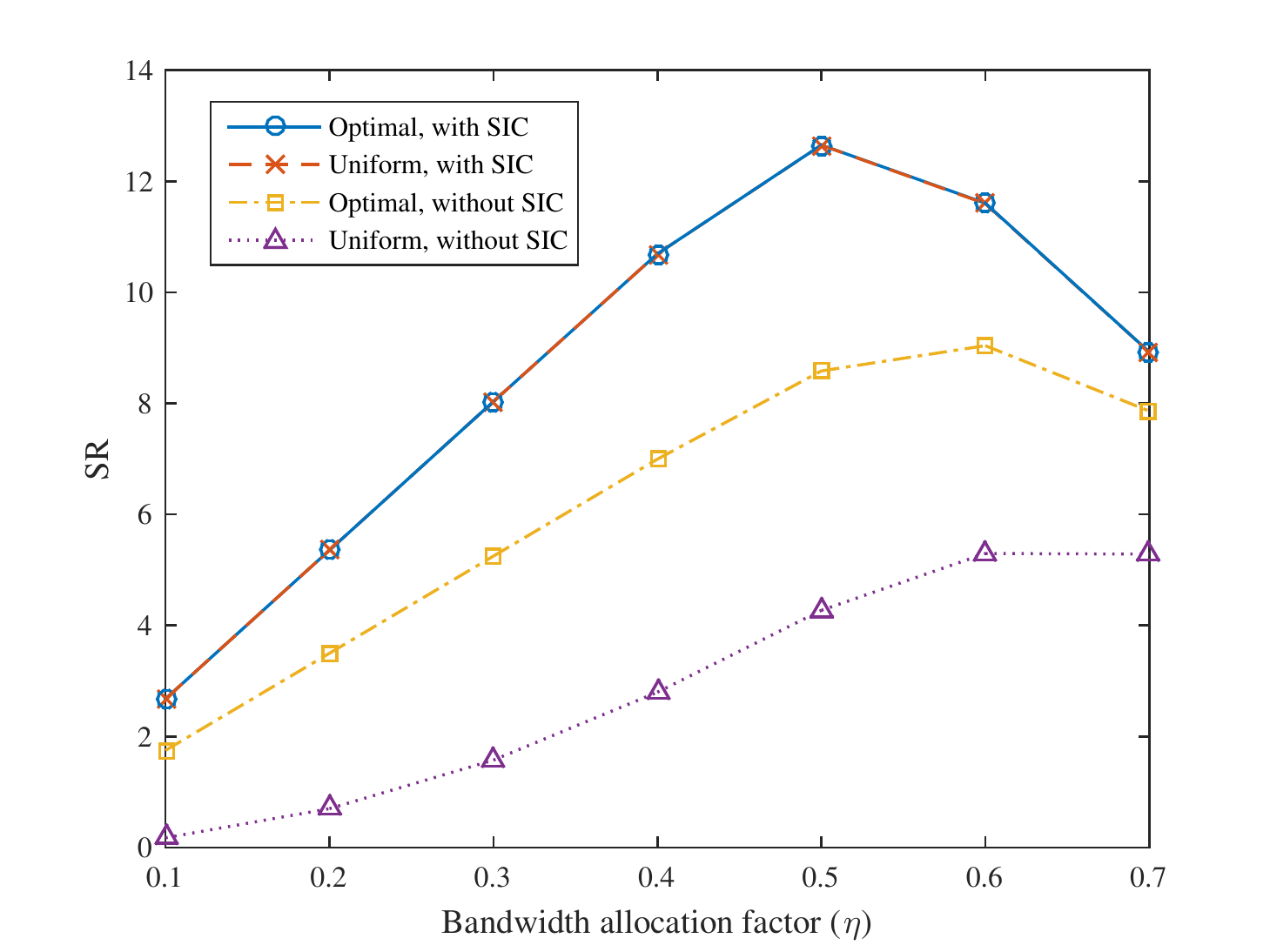}}
  \hspace{1in}
  \subfigure[]{
    \label{SR_versus_eta_p2p}
    \includegraphics[width=0.6\textwidth]{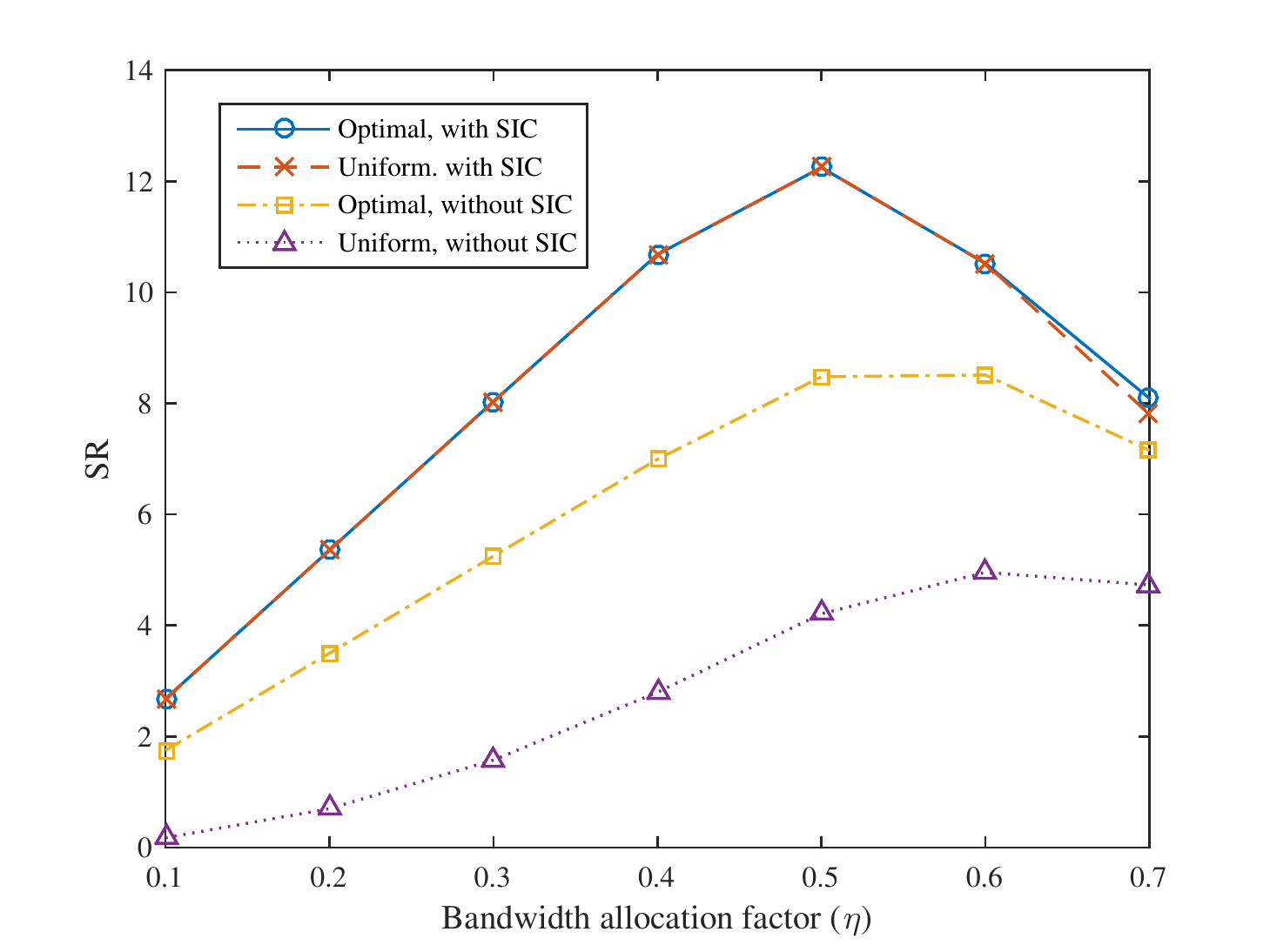}}
  \caption{Comparison of the optimal compression and uniform compression under (a) WZ coding and (b) P2P compression, respectively, with $\{C_0=30, L=4, J=4, \text{and} \ K=2\}$.}
  \label{SR_versus_eta_p2p_wz} %% label for entire figure
\end{figure}

Figs. \ref{SR_versus_eta_wz} and \ref{SR_versus_eta_p2p} compare the optimal compression and uniform compression versus different $\eta$ values with $\{C_0=30, L=4, J=4, \text{and} \ K=2\}$ under WZ coding and P2P compression, respectively.  From Fig. \ref{SR_versus_eta_wz},  we observe that in the case of WZ coding, when the linear reception with SIC  is adopted, the uniform compression can achieve near optimal performance, compared to the optimal compression. However, under the linear reception without SIC, the optimal compression shows a better performance than the uniform compression. On the other hand, in the case of P2P compression as shown in Fig. \ref{SR_versus_eta_p2p}, even though the  linear reception with SIC  is adopted, the uniform compression achieves the near optimal performance only in the small $\eta$-value regime but has a suboptimal performance in the large $\eta$-value regime.   Such observations prove the analysis in Section \ref{subanalysis}, i.e., the approximation in \eqref{appro deter of psi} holds in the condition of high SQNR.
Besides, Figs. \ref{SR_versus_eta_wz} and \ref{SR_versus_eta_p2p} also suggest that the linear reception with SIC always achieves a higher SR than that without SIC under both two cases of the optimal compression and the uniform compression.

\section{Conclusion}\label{conclusion}
In this work, we considered an economic solution to the challenge of limited fronthaul capacity, where the fronthaul links shared  spectrum resource with radio access networks.  To maximize the achievable SR in an uplink H-CRAN, we  jointly optimized fronthaul compression and bandwidth allocation considering different decoding strategies and different compression schemes. Large dimensional random matrix theory was used to derive the deterministic expressions for the ergodic SR. Then, an approximation problem of the joint optimization problem was formulated, which  only depended on statistical channel information. Thus the communication overhead for obtaining channel information and computational complexity were reduced significantly. The simulations results showed that WZ coding always achieved a larger SR than P2P compression at the cost of computational complexity. The linear reception with SIC had a better performance than that without SIC because the former used interference cancellation technique to improve the decoding performance, but the latter decoded each UE's message independently. Actually, to some extent, we could approximately take the achievable SRs under the two decoding strategies as the upper bound and lower bound. Besides, we also provided the performance analysis of uniform compression. Typically, when the linear reception with SIC was applied, the uniform compression with WZ coding could achieve the near optimal performance. Finally, the influence of the normalized fronthaul capacity on the optimal bandwidth allocation coefficient was analyzed. If the normalized fronthaul capacity is much smaller, almost all  the  bandwidth was allocated to the radio access networks. However, in the large normalized fronthaul capacity regime, the optimal bandwidth allocation coefficient, as well as the maximal achievable SR, increased with the normalized fronthaul capacity.

\begin{appendices}
\section{Proof of \textbf{Lemma} 1}\label{appendixA}
We first assume that $R^{(B,SIC)}(\sigma^2)$ is a function of $\sigma^2$. Since $R^{(B,SIC)}(\sigma^2)=\eta R^{(B,SIC)}_{0}(\sigma^2)$ and $\bar{R}^{(B,SIC)}(\sigma^2)=\eta \bar{R}^{(B,SIC)}_{0}(\sigma^2)$, it is equivalent to validating that $R^{(B,SIC)}_{0}(\sigma^2)-\bar{R}^{(B,SIC)}_{0}(\sigma^2)\rightarrow 0$, as $N\rightarrow\infty$.   The derivative of $R^{(B,SIC)}_0(\sigma^2)$ with respect to $\sigma^2$ is expressed as
\begin{align}
  \frac{\partial{R^{(B,SIC)}_0(\sigma^2)}}{\partial{\sigma^2}}=\frac{1}{\log2}\left\{\Ex\left[\tr(\qH\qP^{(M)}\qH^H+\sigma^2\qI_N)^{-1}\right]-\frac{N}{\sigma^2}\right\}.
\end{align}
The relation between the mutual information $R^{(B,SIC)}_0(\sigma^2)$ and $\tr(\qH\qP^{(M)}\qH^H+\sigma^2\qI_N)^{-1}$ can be equivalently written as \cite{hachem2007deterministic,wen2013adetermini},
\begin{align}
 R^{(B,SIC)}_0(\sigma^2)&=\frac{1}{\log2}\int_{-\infty}^{\sigma^2}\left\{\Ex\left[\tr(\qH\qP^{(M)}\qH^H+\omega\qI_N)^{-1}\right]-\frac{N}{\omega}\right\}d\omega.
\end{align}
According to the random matrix theory \cite{zhang2013large}, we get
\begin{align}
  \tr(\qH\qP^{(M)}\qH^H+\omega\qI_N)^{-1}&\asymp\tr\left(\frac{1}{N}\sum\limits_{k=1}^{K}{\frac{Np^{(M)}_{k}\nu_k\qI_N}{1+e_{k}}}+\omega\qI_N\right)^{-1}\\
  &=\tr\left(\bm{\Gamma}^{-1}\right),
\end{align}
where $a\asymp b$ denotes that $a-b\rightarrow 0$ as $N\rightarrow \infty$. Then, we define $\bm{\Sigma}_k=p^{(M)}_{k}\nu_k\qI_N$ and
\begin{equation}
  R(\sigma^2,\qx_k)=\frac{1}{\log2}\left[\log|\sum\limits_{k=1}^K{x_k\Sigma_k}+\sigma^2\qI_N|+\sum\limits_{k=1}^K{(x_k-\log x_k)} -N\log\sigma^2\right],
\end{equation}
where $\qx_k=[x_1,\ldots,x_K]^T$ is a vector consisting of $K$ variables. The derivative of $R^{(B,SIC)}_0(\sigma^2)$ with respect to $\sigma^2$ can be expressed as
\begin{equation}
  \frac{\partial R^{(B,SIC)}_0(\sigma^2)}{\partial \sigma^2}=\frac{\partial R(\sigma^2,\qx_k)}{\partial \sigma^2}|_{x_k=\frac{1}{1+e_k}}+\sum\limits_{k=1}^K{\frac{\partial R(\sigma^2,\qx_k)}{\partial x_k}|_{x_k=\frac{1}{1+e_k}} \frac{\partial e_k}{\partial \sigma^2}}.
\end{equation}
Note that $\frac{\partial R(\sigma^2,\qx_k)}{\partial x_k}|_{x_k=\frac{1}{1+e_k}}=0$,  therefore
\begin{equation}
\frac{\partial R^{(B,SIC)}_0(\sigma^2)}{\partial \sigma^2}=\frac{\partial R}{\partial \sigma^2}|_{x_k=\frac{1}{1+e_k}},
\end{equation}
which is equal to \eqref{asym rp}.

Taking the same actions on  the numerator and denominator of $R^{(B,L)}_0(\eta)$, respectively, then we have
\begin{equation}\label{eeq1}
 \log_2|\qH\qP^{(M)}\qH^H+\sigma^2\qI_N|-\frac{1}{\log_2}\Delta_0\rightarrow0,
\end{equation}
and
\begin{equation}\label{eeq2}
 \log_2|\qH_{[k]}\qP^{(M)}_{[k]}\qH^H_{[k]}+\sigma^2\qI_N|-\frac{1}{\log_2}\Delta_k \rightarrow0.
\end{equation}
Finally, the results of \eqref{eeq1} and \eqref{eeq2} can be formed into $\bar{R}^{(B,L)}_0(\eta)$. Hence the proof is completed.$\hfill\blacksquare$
\section{Proof of \textbf{Proposition 1}}\label{appendixB}
The proof of \textbf{Proposition 1}  is based on \cite{zappone2015energy}. We assume $\omega^{\ast}$ is the optimal solution to \eqref{joint op of asym2 problem}, then
\begin{equation}
 \omega^{\ast}=\frac{f(\bm{\Psi}^{\ast},\bm{\Omega}^{\ast})}{g(\bm{\Psi}^{\ast})}\geq\frac{f(\bm{\Psi},\bm{\Omega})}{g(\bm{\Psi})},
\end{equation}
which implies
\begin{equation}
  f(\bm{\Psi},\bm{\Omega})-\omega^{\ast}g(\bm{\Psi})\leq0,
\end{equation}
 with the equality if and only if $\bm{\Psi}=\bm{\Psi}^{\ast}$ and $\bm{\Omega}=(\bm{\Psi}^{\ast}+\sigma^2\qI_L)^{-1}$.

 On the other side, it is assumed that $(\bm{\Psi}^{\ast},\bm{\Omega}^{\ast})=\mathop{\argmax}_{\bm{\Psi}\in\mathcal{S}}[f(\bm{\Psi},\bm{\Omega})-\omega^{\ast}g(\bm{\Psi})]$, suggesting
 \begin{equation}
  f(\bm{\Psi},\bm{\Omega})-\omega^{\ast}g(\bm{\Psi})\leq f(\bm{\Psi}^{\ast},\bm{\Omega}^{\ast})-\omega^{\ast}g(\bm{\Psi}^{\ast})=0,\forall \bm{\Psi}\in \mathcal{S} \  \text{and} \ \forall \bm{\Omega},
 \end{equation}
 which in turn be rewritten as the following conditions:
 \begin{equation}
   \omega^{\ast}\geq\frac{ f(\bm{\Psi},\bm{\Omega})}{g(\bm{\Psi})}, \forall \bm{\Psi}\in\mathcal{S} \ \text{and} \ \forall \bm{\Omega},
 \end{equation}
 and
 \begin{equation}
   \omega^{\ast}=\frac{ f(\bm{\Psi}^{\ast},\bm{\Omega}^{\ast})}{g(\bm{\Psi}^{\ast})}.
 \end{equation}
 Therefore, the proof is completed. $\hfill\blacksquare$

\section{Proof of \textbf{Proposition 2}}\label{AppendixC}
 The update rule for $\omega$ can be written as
 \begin{equation}\label{update rule}
   \omega^{(t+1)}=\frac{f(\bm{\Psi}^{(t)},\bm{\Omega}^{(t)})}{g(\bm{\Psi}^{(t)})}= \omega^{(t)}-\frac{f(\bm{\Psi}^{(t)},\bm{\Omega}^{(t)})-\omega^{(t)}g(\bm{\Psi}^{(t)})}{-g(\bm{\Psi}^{(t)})},
 \end{equation}
then
\begin{equation}
  F(\omega^{(t)})=(\omega^{(t+1)}-\omega^{(t)})g(\bm{\Psi}^{(t)}).
\end{equation}
According to \cite[Lemma 3.1-d]{zappone2015energy}, we have $F(\omega^{(t)})\geq 0$. Since function $g(\bm{\Psi}^{(t)})>0$, $\omega^{(t+1)}\geq\omega^{(t)}$ with the equality if and only if we have already been at convergence. Therefore, the sequence of $\{\omega^{(t)}\}$ is an increasing sequence, leading to a decreasing succession $\{F(\omega^{(t)})\}$. Given $F(\omega)$ is continuous over compact domain  $\mathcal{S}$, \textbf{Algorithm 2} converges to an optimal solution. $\hfill\blacksquare$
\end{appendices}

\bibliographystyle{IEEEtran}
\bibliography{reference}

\begin{thebibliography}{10}
\providecommand{\url}[1]{#1}
\csname url@samestyle\endcsname
\providecommand{\newblock}{\relax}
\providecommand{\bibinfo}[2]{#2}
\providecommand{\BIBentrySTDinterwordspacing}{\spaceskip=0pt\relax}
\providecommand{\BIBentryALTinterwordstretchfactor}{4}
\providecommand{\BIBentryALTinterwordspacing}{\spaceskip=\fontdimen2\font plus
\BIBentryALTinterwordstretchfactor\fontdimen3\font minus
  \fontdimen4\font\relax}
\providecommand{\BIBforeignlanguage}[2]{{%
\expandafter\ifx\csname l@#1\endcsname\relax
\typeout{** WARNING: IEEEtran.bst: No hyphenation pattern has been}%
\typeout{** loaded for the language `#1'. Using the pattern for}%
\typeout{** the default language instead.}%
\else
\language=\csname l@#1\endcsname
\fi
#2}}
\providecommand{\BIBdecl}{\relax}
\BIBdecl

\bibitem{Xia2017Joint}
W.~Xia, J.~Zhang, T.~Q.~S. Quek, S.~Jin, and H.~Zhu, ``Joint optimization of
  fronthaul compression and bandwidth allocation in heterogeneous {CRAN},'' in
  \emph{Proc. IEEE Global Commun. Conf. (GLOBECOM)}, Singapore, Dec. 2017, pp.
  1--6.

\bibitem{Peng2015System}
M.~Peng, Y.~Li, Z.~Zhao, and C.~Wang, ``System architecture and key
  technologies for {5G} heterogeneous cloud radio access networks,'' \emph{IEEE
  Network}, vol.~29, no.~2, pp. 6--14, Mar. 2015.

\bibitem{tony2017cloud}
T.~Q.~S. Quek, M.~Peng, O.~Simeone, and W.~Yu, \emph{Cloud Radio Access
  Networks: Principles, Technologies, and Applications}.\hskip 1em plus 0.5em
  minus 0.4em\relax Cambridge Univ. Press, 2017.

\bibitem{Gao2018throyghput}
H.~Gao, J.~Gao, Z.~Shi, and T.~Lv, ``Throughput and energy efficiency of
  wireless powered multi-tier {MIMO} {HetNets},'' \emph{J. Signal Process.
  Systems}, vol.~90, no.~6, pp. 857--871, Jun. 2018.

\bibitem{Tang2017System}
J.~Tang, W.~P. Tay, T.~Q.~S. Quek, and B.~Liang, ``System cost minimization in
  cloud {RAN} with limited fronthaul capacity,'' \emph{IEEE Trans. Wireless
  Commun.}, vol.~16, no.~5, pp. 3371--3384, May 2017.

\bibitem{Peng2015Contract}
M.~Peng, X.~Xie, Q.~Hu, J.~Zhang, and H.~V. Poor, ``Contract-based interference
  coordination in heterogeneous cloud radio access networks,'' \emph{IEEE J.
  Sel. Areas Commun.}, vol.~33, no.~6, pp. 1140--1153, Jun. 2015.

\bibitem{peng2014heterogeneous}
M.~Peng, Y.~Li, J.~Jiang, J.~Li, and C.~Wang, ``Heterogeneous cloud radio
  access networks: A new perspective for enhancing spectral and energy
  efficiencies,'' \emph{IEEE Wireless Commun.}, vol.~21, no.~6, pp. 126--135,
  Dec. 2014.

\bibitem{tang2015cross}
J.~Tang, W.~P. Tay, and T.~Q.~S. Quek, ``Cross-layer resource allocation with
  elastic service scaling in cloud radio access network,'' \emph{IEEE Trans.
  Wireless Commun.}, vol.~14, no.~9, pp. 5068--5081, Sep. 2015.

\bibitem{Yang2016Energy}
H.~H. Yang, G.~Geraci, and T.~Q.~S. Quek, ``Energy-efficient design of {MIMO}
  heterogeneous networks with wireless backhaul,'' \emph{IEEE Trans. Wireless
  Commun.}, vol.~15, no.~7, pp. 4914--4927, Jul. 2016.

\bibitem{zhang2017downlink}
H.~Zhang, H.~Liu, J.~Cheng, and V.~C.~M. Leung, ``Downlink energy efficiency of
  power allocation and wireless backhaul bandwidth allocation in heterogeneous
  small cell networks,'' \emph{IEEE Trans. Commun.}, vol.~66, no.~4, pp.
  1705--1716, Apr. 2017.

\bibitem{Nguyen2016Resource}
T.~M. Nguyen, A.~Yadav, W.~Ajib, and C.~Assi, ``Resource allocation in two-tier
  wireless backhaul heterogeneous networks,'' \emph{IEEE Trans. Wireless
  Commun.}, vol.~15, no.~10, pp. 6690--6704, Oct. 2016.

\bibitem{Wang2016Joint}
N.~Wang, E.~Hossain, and V.~K. Bhargava, ``Joint downlink cell association and
  bandwidth allocation for wireless backhauling in two-tier {HetNets} with
  large-scale antenna arrays,'' \emph{IEEE Trans. Wireless Commun.}, vol.~15,
  no.~5, pp. 3251--3268, May 2016.

\bibitem{Xia2016Bandwidth}
W.~Xia, J.~Zhang, S.~Jin, C.-K. Wen, F.~Gao, and H.~Zhu, ``Bandwidth allocation
  in heterogeneous networks with wireless backhaul,'' in \emph{Proc. IEEE
  Global Commun. Conf. (GLOBECOM)}, Washington DC, USA, Dec. 2016, pp. 1--6.

\bibitem{park2014performance}
S.-H. Park, O.~Simeone, O.~Sahin, and S.~Shamai, ``Performance evaluation of
  multiterminal backhaul compression for cloud radio access networks,'' in
  \emph{Proc. CISS}, Princeton, NJ, USA, Mar. 2014, pp. 1--6.

\bibitem{park2014fronthaul}
S.-H. Park, O.~Simeone, O.~Sahin, and S.~S. Shitz, ``Fronthaul compression for
  cloud radio access networks: Signal processing advances inspired by network
  information theory,'' \emph{IEEE Signal Process. Mag.}, vol.~31, no.~6, pp.
  69--79, Nov. 2014.

\bibitem{zhou2014optimized}
Y.~Zhou and W.~Yu, ``Optimized backhaul compression for uplink cloud radio
  access network,'' \emph{IEEE J. Sel. Areas Commun.}, vol.~32, no.~6, pp.
  1295--1307, Jun. 2014.

\bibitem{zhou2016fronthaul}
------, ``Fronthaul compression and transmit beamforming optimization for
  multi-antenna uplink {C-RAN},'' \emph{IEEE Trans. Signal Process.}, vol.~64,
  no.~16, pp. 4138--4151, Aug. 2016.

\bibitem{park2013robust}
S.-H. Park, O.~Simeone, O.~Sahin, and S.~Shamai, ``Robust and efficient
  distributed compression for cloud radio access networks,'' \emph{IEEE Trans.
  Veh. Technol.}, vol.~62, no.~2, pp. 692--703, Feb. 2013.

\bibitem{vu2017adaptive}
T.~X. Vu, H.~D. Nguyen, T.~Q.~S. Quek, and S.~Sun, ``Adaptive cloud radio
  access networks: Compression and optimization,'' \emph{IEEE Trans. Signal
  Process.}, vol.~65, no.~1, pp. 228--241, Jan. 2017.

\bibitem{Park2013Joint}
S.-H. Park, O.~Simeone, O.~Sahin, and S.~Shamai, ``Joint precoding and
  multivariate backhaul compression for the downlink of cloud radio access
  networks,'' \emph{IEEE Trans. Signal Process.}, vol.~61, no.~22, pp.
  5646--5658, Nov. 2013.

\bibitem{Park2014Inter}
------, ``Inter-cluster design of precoding and fronthaul compression for cloud
  radio access networks,'' \emph{IEEE Wireless Commun. Lett.}, vol.~3, no.~4,
  pp. 369--372, Aug. 2014.

\bibitem{simeone2016cloud}
O.~Simeone, A.~Maeder, M.~Peng, O.~Sahin, and W.~Yu, ``Cloud radio access
  network: Virtualizing wireless access for dense heterogeneous systems,''
  \emph{J. Commun. Netw.}, vol.~18, no.~2, pp. 135--149, Apr. 2016.

\bibitem{chi2017message}
Y.~Chi, L.~Liu, G.~Song, C.~Yuen, Y.~L. Guan, and Y.~Li, ``Message passing in
  {C-RAN}: {Joint} user activity and signal detection,'' in \emph{Proc. IEEE
  Global Commun. Conf. (GLOBECOM)}, Singapore, Dec. 2017.

\bibitem{Liu2016Gaussian}
L.~Liu, C.~Yuen, Y.~L. Guan, Y.~Li, and C.~Huang, ``Gaussian message passing
  iterative detection for {MIMO-NOMA} systems with massive access,'' in
  \emph{Proc. IEEE Global Commun. Conf. (GLOBECOM)}, Washington DC, USA, Dec.
  2016, pp. 1--6.

\bibitem{zhou2013approximate}
Y.~Zhou and W.~Yu, ``Approximate bounds for limited backhaul uplink multicell
  processing with single-user compression,'' in \emph{Proc. IEEE Canadian
  Workshop Inf. Theory (CWIT)}, Toronto, ON, Canada, Jun. 2013, pp. 113--116.

\bibitem{zhang2014the}
X.~Zhang and M.~Haenggi, ``The performance of successive interference
  cancellation in random wireless networks,'' \emph{IEEE Trans. Inf. Theory},
  vol.~60, no.~10, pp. 6368--6388, Oct. 2014.

\bibitem{zhang2017energy}
H.~Zhang, B.~Wang, C.~Jiang, K.~Long, A.~Nallanathan, and V.~C.~M. Leung,
  ``Energy efficient dynamic resource allocation in {NOMA} networks,'' in
  \emph{Proc. IEEE Global Commun. Conf. (GLOBECOM)}, Singapore, Dec. 2017, pp.
  1--5.

\bibitem{lin2017anew}
H.~Lin, F.~Gao, S.~Jin, and G.~Y. Li, ``A new view of multi-user hybrid massive
  {MIMO}: {Non}-orthogonal angle division multiple access,'' \emph{IEEE J. Sel.
  Areas Commun.}, vol.~35, no.~10, pp. 2268--2280, Oct. 2017.

\bibitem{beck2009gradient}
A.~Beck and M.~Teboulle, ``Gradient-based algorithms with applications to
  signal recovery,'' \emph{Convex Optimization Signal Process. Commun.}, pp.
  42--88, 2009.

\bibitem{zhang2013large}
J.~Zhang, C.-K. Wen, S.~Jin, X.~Gao, and K.-K. Wong, ``Large system analysis of
  cooperative multi-cell downlink transmission via regularized channel
  inversion with imperfect {CSIT},'' \emph{IEEE Trans. Wireless Commun.},
  vol.~12, no.~10, pp. 4801--4813, Oct. 2013.

\bibitem{xia2017large}
W.~Xia, J.~Zhang, S.~Jin, C.~K. Wen, F.~Gao, and H.~Zhu, ``Large system
  analysis of resource allocation in heterogeneous networks with wireless
  backhaul,'' \emph{IEEE Trans. Commun.}, vol.~65, no.~11, pp. 5040--5053, Nov.
  2017.

\bibitem{wen2013adetermini}
C.-K. Wen, G.~Pan, K.-K. Wong, M.~Guo, and J.~C. Chen, ``A deterministic
  equivalent for the analysis of non-gaussian correlated {MIMO} multiple access
  channels,'' \emph{IEEE Trans. Inf. Theory}, vol.~59, no.~1, pp. 329--352,
  Jan. 2013.

\bibitem{zhang2013oncap}
J.~Zhang, C.-K. Wen, S.~Jin, X.~Gao, and K.-K. Wong, ``On capacity of
  large-scale {MIMO} multiple access channels with distributed sets of
  correlated antennas,'' \emph{IEEE J. Sel. Areas Commun.}, vol.~31, no.~2, pp.
  133--148, Feb. 2013.

\bibitem{dinkelbach1967nonlinear}
W.~Dinkelbach, ``On nonlinear fractional programming,'' \emph{Management
  science}, vol.~13, no.~7, pp. 492--498, Mar. 1967.

\bibitem{zappone2015energy}
A.~Zappone, E.~Jorswieck \emph{et~al.}, ``Energy efficiency in wireless
  networks via fractional programming theory,'' \emph{Found. Trends Commun.
  Inf. Theory}, vol.~11, no. 3-4, pp. 185--396, 2015.

\bibitem{li2013transmit}
Q.~Li, M.~Hong, H.-T. Wai, Y.-F. Liu, W.-K. Ma, and Z.-Q. Luo, ``Transmit
  solutions for {MIMO} wiretap channels using alternating optimization,''
  \emph{IEEE J. Sel. Areas Commun.}, vol.~31, no.~9, pp. 1714--1727, Sep. 2013.

\bibitem{GRIPPO200On}
L.~Grippo and M.~Sciandrone, ``On the convergence of the block nonlinear
  gauss¨cseidel method under convex constraints,'' \emph{Oper. Res. Lett.},
  vol.~26, no.~3, pp. 127 -- 136, Apr. 2000.

\bibitem{Peng2015Energy}
M.~Peng, K.~Zhang, J.~Jiang, J.~Wang, and W.~Wang, ``Energy-efficient resource
  assignment and power allocation in heterogeneous cloud radio access
  networks,'' \emph{IEEE Trans. Veh. Technol.}, vol.~64, no.~11, pp.
  5275--5287, Nov. 2015.

\bibitem{hachem2007deterministic}
W.~Hachem, P.~Loubaton, J.~Najim \emph{et~al.}, ``Deterministic equivalents for
  certain functionals of large random matrices,'' \emph{Ann. Appl. Probab.},
  vol.~17, no.~3, pp. 875--930, 2007.

\end{thebibliography}

\end{document}